\documentclass[11pt]{article}
\usepackage{amssymb,graphicx,multicol}
\usepackage{times}
\usepackage{psfrag}
\skip\footins 39pt plus 4pt minus 2pt
\def\footnoterule{\kern-19pt\hrule width.5in\kern18.6pt}%
\newcommand{\dotsb}{\ldots}

\newcommand{\half}{\frac{1}{2}}
\newcommand{\ts}{\hskip0.1ex\raisebox{-1ex}[0ex][0.8ex]{\rule{0.1ex}{2.75ex}\hskip0.2ex}}

\newcommand{\Caption}[1]{\caption{\small #1}}
\begin{document}
%
\newcommand{\fig}[2]{\includegraphics[width=#1]{./figures/#2}}
\newcommand{\Fig}[1]{\includegraphics[width=\columnwidth]{./figures/#1}}
\newlength{\bilderlength}
\newcommand{\bilderscale}{0.35}
\newcommand{\storebilderscale}{\bilderscale}
\newcommand{\bilderskip}{\hspace*{0.8ex}}
\newcommand{\textdiagram}[1]{%
\renewcommand{\bilderscale}{0.2}%
\diagram{#1}\renewcommand{\bilderscale}{\storebilderscale}}
\newcommand{\vardiagram}[2]{%
\renewcommand{\bilderscale}{#1}%
\diagram{#2}\renewcommand{\bilderscale}{\storebilderscale}}
\newcommand{\diagram}[1]{%
\settowidth{\bilderlength}{\bilderskip%
\includegraphics[scale=\bilderscale]{./figures/#1}\bilderskip}%
\parbox{\bilderlength}{\bilderskip%
\includegraphics[scale=\bilderscale]{./figures/#1}\bilderskip}}
\newcommand{\Diagram}[1]{%
\settowidth{\bilderlength}{%
\includegraphics[scale=\bilderscale]{./figures/#1}}%
\parbox{\bilderlength}{%
\includegraphics[scale=\bilderscale]{./figures/#1}}}
%

%
\newcommand{\sgn}{{\mathrm{sgn}}}
\newcommand{\rme}{{\mathrm{e}}}
\newcommand{\rmd}{{\mathrm{d}}}
\newcommand{\nn}{\nonumber}\newcommand {\eq}[1]{(\ref{#1})}
\newcommand {\Eq}[1]{Eq.\hspace{0.55ex}(\ref{#1})}
\newcommand {\Eqs}[1]{Eqs.\hspace{0.55ex}(\ref{#1})}
\newcommand{\E}{\epsilon}
\newcommand{\R}{\mathbb{R}}
\newcommand{\N}{\mathbb{N}}

\def\true{true}
\newsavebox{\bilderbox}
\newlength{\bilderhelp}
\newsavebox{\bilderone}
\newlength{\bilderonelength}
\newsavebox{\bildertwo}
\newlength{\bildertwolength}
%
\newcommand{\bild}[1]{\fboxsep0mm%
\sbox{\bilderbox}{
{\includegraphics[scale=\bilderscale]{#1}}}%
\settowidth{\bilderlength}{\usebox{\bilderbox}}%
\parbox{\bilderlength}{\usebox{\bilderbox}}}
\newcommand{\savebild}[3]{\newsavebox{#2}%
\sbox{#2}{\bild{#3}}\newcommand{#1}{%
\ensuremath{\,\mathchoice{\usebox{#2}}%
{\settowidth{\bilderhelp}{\scalebox{0.7}{\usebox{#2}}}%
\parbox{\bilderhelp}{\scalebox{0.7}{\usebox{#2}}}}%
{\settowidth{\bilderhelp}{\scalebox{0.5}{\usebox{#2}}}%
\parbox{\bilderhelp}{\scalebox{0.5}{\usebox{#2}}}}%
{\settowidth{\bilderhelp}{\scalebox{0.35}{\usebox{#2}}}%
\parbox{\bilderhelp}{\scalebox{0.35}{\usebox{#2}}}}%
\,}}}
\newcommand{\bilderdiagram}[4]{{%
\mathchoice{
\sbox{\bilderone}{\ensuremath{\displaystyle#1}}%
\sbox{\bildertwo}{\ensuremath{\displaystyle#2}}%
\settoheight{\bilderonelength}{\ensuremath{\usebox{\bilderone}}}%
\settoheight{\bildertwolength}{\ensuremath{\usebox{\bildertwo}}}%
\left#3\!\usebox{\bilderone}{\rule{0mm}{\bildertwolength}}\right.%
\hspace*{-0.5ex}\!\left|\!{\rule{0mm}{\bilderonelength}}%
\usebox{\bildertwo}\!\right#4}%
{
\sbox{\bilderone}{\ensuremath{\textstyle#1}}%
\sbox{\bildertwo}{\ensuremath{\textstyle#2}}%
\settoheight{\bilderonelength}{\ensuremath{\usebox{\bilderone}}}%
\settoheight{\bildertwolength}{\ensuremath{\usebox{\bildertwo}}}%
\left#3\!\usebox{\bilderone}{\rule{0mm}{\bildertwolength}}\right.%
\hspace*{-0.35ex}\!\left|\!{\rule{0mm}{\bilderonelength}}%
\usebox{\bildertwo}\!\right#4}%
{
\sbox{\bilderone}{\ensuremath{\scriptstyle#1}}%
\sbox{\bildertwo}{\ensuremath{\scriptstyle#2}}%
\settoheight{\bilderonelength}{\ensuremath{\usebox{\bilderone}}}%
\settoheight{\bildertwolength}{\ensuremath{\usebox{\bildertwo}}}%
\left#3\!\usebox{\bilderone}{\rule{0mm}{\bildertwolength}}\right.%
\hspace*{-0.1ex}\!\left|\!{\rule{0mm}{\bilderonelength}}%
\usebox{\bildertwo}\!\right#4}%
{
\sbox{\bilderone}{\ensuremath{\scriptscriptstyle#1}}%
\sbox{\bildertwo}{\ensuremath{\scriptscriptstyle#2}}%
\settoheight{\bilderonelength}{\ensuremath{\usebox{\bilderone}}}%
\settoheight{\bildertwolength}{\ensuremath{\usebox{\bildertwo}}}%
\left#3\!\usebox{\bilderone}{\rule{0mm}{\bildertwolength}}\right.%
\hspace*{-0.1ex}\!\left|\!{\rule{0mm}{\bilderonelength}}%
\usebox{\bildertwo}\!\right#4}%
}}
\newcommand{\MOPE}[2]{\bilderdiagram{#1}{#2}{(}{)}}
\newcommand{\DIAG}[2]{\bilderdiagram{#1}{#2}{<}{>}}
\newcommand{\DIAGindhelp}[4]{%
\sbox{\bilderbox}{\ensuremath{#4\bilderdiagram{#1}{#2}{<}{>}}}%
\settowidth{\bilderlength}{\rotatebox{90}{\ensuremath{\usebox{\bilderbox}}}}%
\ensuremath{\usebox{\bilderbox}_{\hspace*{-0.162\bilderlength}#3}}}
\newcommand{\DIAGind}[3]{%
\mathchoice{\DIAGindhelp{#1}{#2}{#3}{\displaystyle}}%
{\DIAGindhelp{#1}{#2}{#3}{\textstyle}}%
{\DIAGindhelp{#1}{#2}{#3}{\scriptstyle}}%
{\DIAGindhelp{#1}{#2}{#3}{\scriptscriptstyle}}}
\newcommand{\reducedbildheightrule}[2]{{%
\mathchoice{\settowidth{\bilderlength}{\rotatebox{90}{\ensuremath{\displaystyle#1}}}%
\parbox{0mm}{\rule{0mm}{#2\bilderlength}}}%
{\settowidth{\bilderlength}{\rotatebox{90}{\ensuremath{\textstyle#1}}}%
\parbox{0mm}{\rule{0mm}{#2\bilderlength}}}%
{\settowidth{\bilderlength}{\rotatebox{90}{\ensuremath{\scriptstyle#1}}}%
\parbox{0mm}{\rule{0mm}{#2\bilderlength}}}%
{\settowidth{\bilderlength}{\rotatebox{90}{\ensuremath{\scriptscriptstyle#1}}}%
\parbox{0mm}{\rule{0mm}{#2\bilderlength}}}}}
\newcommand{\bildheightrule}[1]{\reducedbildheightrule{#1}{1}}
\newcommand{\reducedbild}[2]{%
{\settowidth{\bilderhelp}{#1}%
\setlength{\bilderhelp}{#2\bilderhelp}%
\parbox{\bilderhelp}{\scalebox{#2}{#1}}}}
%
%

\centerline{\sffamily\bfseries\large   Functional Renormalization
for Disordered Systems}\smallskip

\centerline{\sffamily\bfseries\large
Basic Recipes and Gourmet Dishes} \smallskip

\bigskip
\centerline{\sffamily\bfseries\normalsize Kay J\"org Wiese and Pierre
Le Doussal}
\bigskip
\centerline{Laboratoire de Physique Th\'eorique de l'Ecole Normale
Superieure, 24 rue Lhomond, 75005 Paris, France.}
\centerline{\it and}
\centerline{KITP, UCSB, Santa Barbara, CA 03106-4030, USA}
\medskip

\medskip
\centerline{\small Nov.~12, 2006}

\noindent \rule{\textwidth}{0.3mm} \smallskip \leftline{\bfseries
Abstract} We give a pedagogical introduction into the functional
renormalization group treatment of disordered systems. After a review
of its phenomenology, we show why in the context of disordered systems
a functional renormalization group treatment is necessary, contrary to
pure systems, where renormalization of a single coupling constant is
sufficient. This leads to a disorder distribution, which after a
finite renormalization becomes non-analytic, thus overcoming the
predictions of the seemingly exact dimensional reduction.  We discuss,
how the non-analyticity can be measured in a simulation or experiment.
We then construct a renormalizable field theory beyond leading
order. We  discuss an elastic manifold embedded in $N$ dimensions,
and give the exact solution for $N\to\infty$. This is compared to
predictions of the Gaussian replica variational ansatz, using replica
symmetry breaking. We further consider random field magnets, and
supersymmetry. We finally discuss depinning, both isotropic and
anisotropic, and universal scaling function.

\noindent\rule{\textwidth}{0.3mm}

\begin{multicols}{2}

{\small\tableofcontents}
\end{multicols}

\noindent\rule{\textwidth}{0.3mm}


\section{Introduction}\label{intro}

Statistical mechanics is by now a rather mature branch of physics.
For pure systems like a ferromagnet, it allows to calculate so precise
details as the behavior of the specific heat on approaching the
Curie-point. We know that it diverges as a function of the distance in
temperature to the Curie-temperature, we know that this divergence has
the form of a power-law, we can calculate the exponent, and we can do
this with at least 3 digits of accuracy. Best of all, these findings
are in excellent agreement with the most precise experiments. This is
a true success story of statistical mechanics.  On the other hand, in
nature no system is really pure, i.e.\ without at least some disorder
(``dirt'').  As experiments (and theory) seem to suggest, a little bit
of disorder does not change the behavior much. Otherwise experiments
on the specific heat of Helium would not so extraordinarily well
confirm theoretical predictions. But what happens for strong disorder?
By this we mean that disorder completely dominates over entropy. Then
already the question: ``What is the ground-state?'' is no longer
simple. This goes hand in hand with the appearance of so-called
metastable states. States, which in energy are very close to the
ground-state, but which in configuration-space may be far apart. Any
relaxational dynamics will take an enormous time to find the correct
ground-state, and may fail altogether, as can be seen in
computer-simulations as well as in experiments. This means that our
way of thinking, taught in the treatment of pure systems, has to be
adapted to account for disorder. We will see that in contrast to pure
systems, whose universal large-scale properties can be described by
very few parameters, disordered systems demand the knowledge of the
whole disorder-distribution function (in contrast to its first few
moments). We show how universality nevertheless emerges.

Experimental realizations of strongly disordered systems are glasses,
or more specifically spin-glasses, vortex-glasses, electron-glasses
and structural glasses (not treated here).  Furthermore random-field
magnets, and last not least elastic systems in disorder.

What is our current understanding of disordered systems? It is here
that the success story of statistical mechanics, with which we started,
comes to an end: Despite 30 years of research, we do not know much:
There are a few exact solutions, there are phenomenological methods
(like the droplet-model), and there is the mean-field approximation,
involving a method called replica-symmetry breaking (RSB). This method
is correct for infinitely connected systems, e.g.\ the SK-model
(Sherrington Kirkpatrick model), or for systems with infinitely many
components.  However it is unclear, to which extend it applies to real
physical systems, in which each degree of freedom is directly coupled only to a
finite number of other degrees of freedom.

Another interesting system are elastic manifolds in a random medium,
which has the advantage of being approachable by other (analytic)
methods, while still retaining all the rich physics of strongly
disordered systems.  Here, we review recent advances.
This review is an extended version of \cite{Wiese2002,Wiese2003a}.
For lectures on the internet see \cite{LeDoussalWindsor2004,LeDoussalKITP2006,WieseKITP2006}.

\section{Physical realizations, model and observables}\label{model}
\begin{figure}[t]
\centerline{\fig{0.25\textwidth}{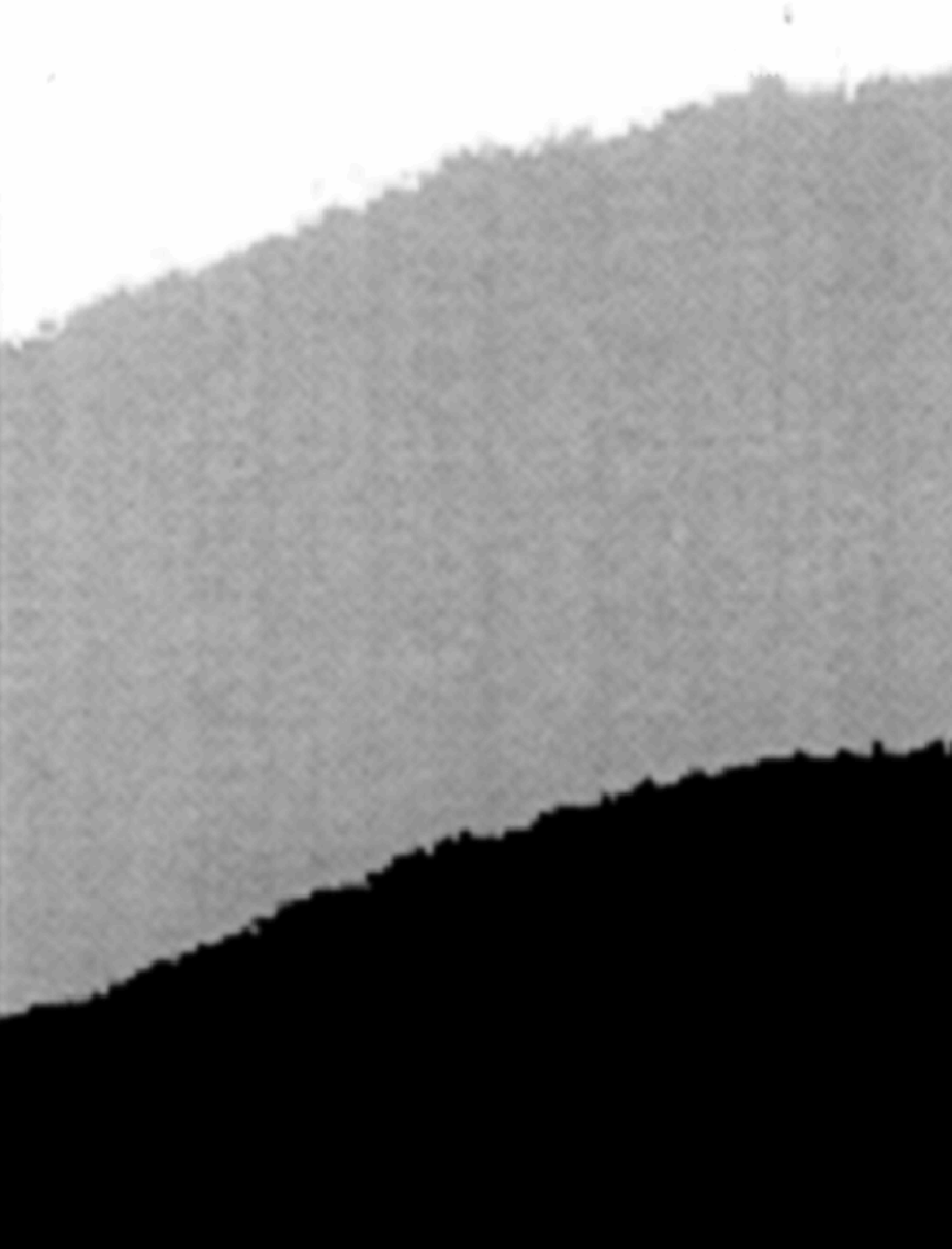}~~~\fig{0.7\textwidth}{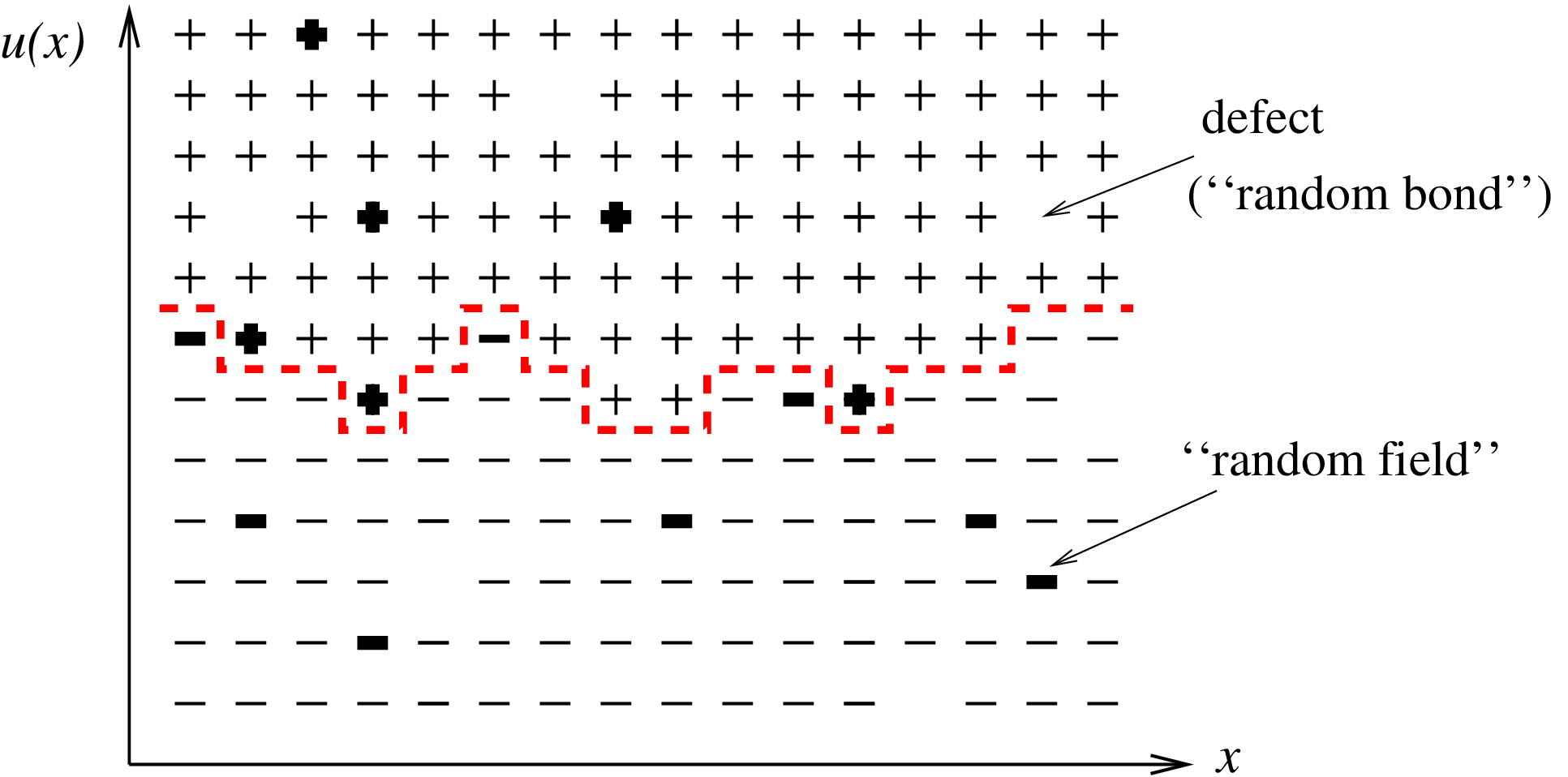}}
\Caption{An Ising magnet at low temperatures forms a domain wall
described by a function $u (x)$ (right). An experiment on a thin
Cobalt film (left)
\protect\cite{LemerleFerreChappertMathetGiamarchiLeDoussal1998}; with
kind permission of the authors.}
\label{exp:Magnet}
\end{figure}
\begin{figure}[b]
\centerline{\parbox{0.5\textwidth}{\fig{0.5\textwidth}{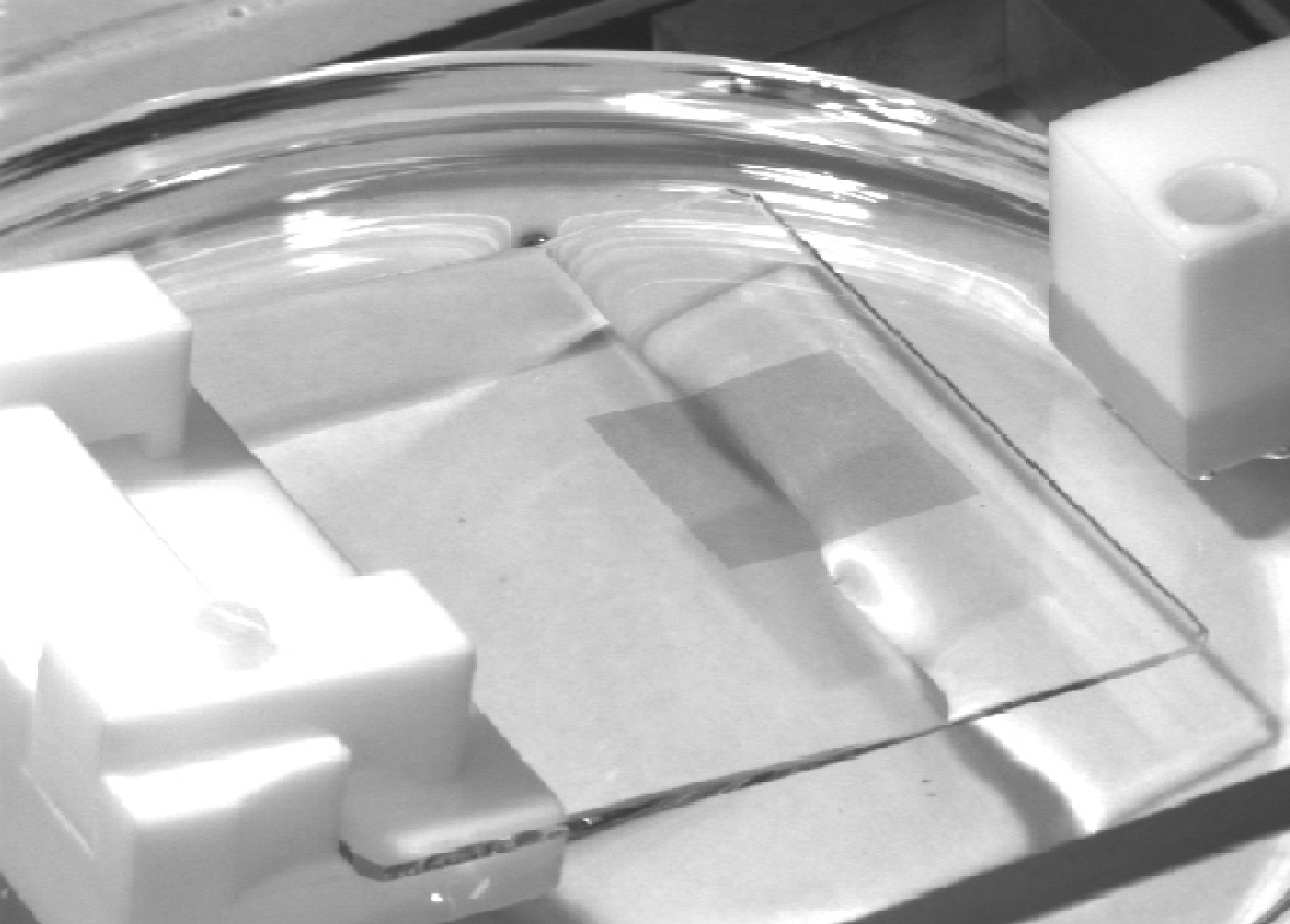}}
\parbox{0.445\textwidth}{\begin{minipage}{0.445\textwidth}
\Fig{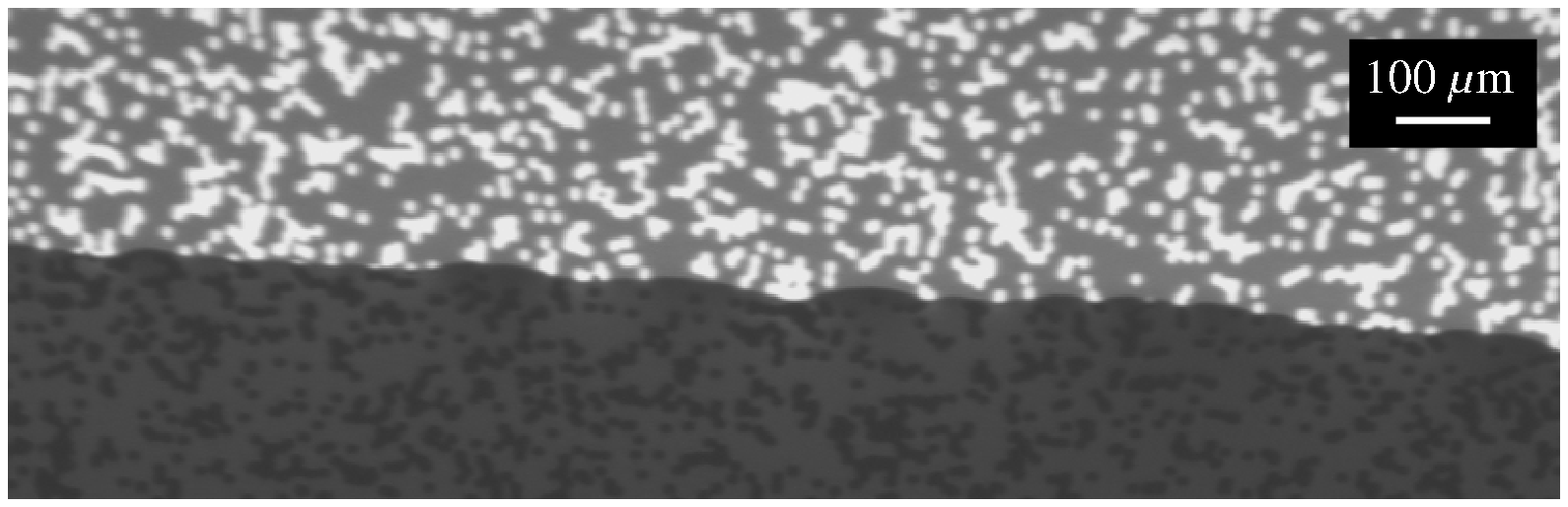}\\
\Fig{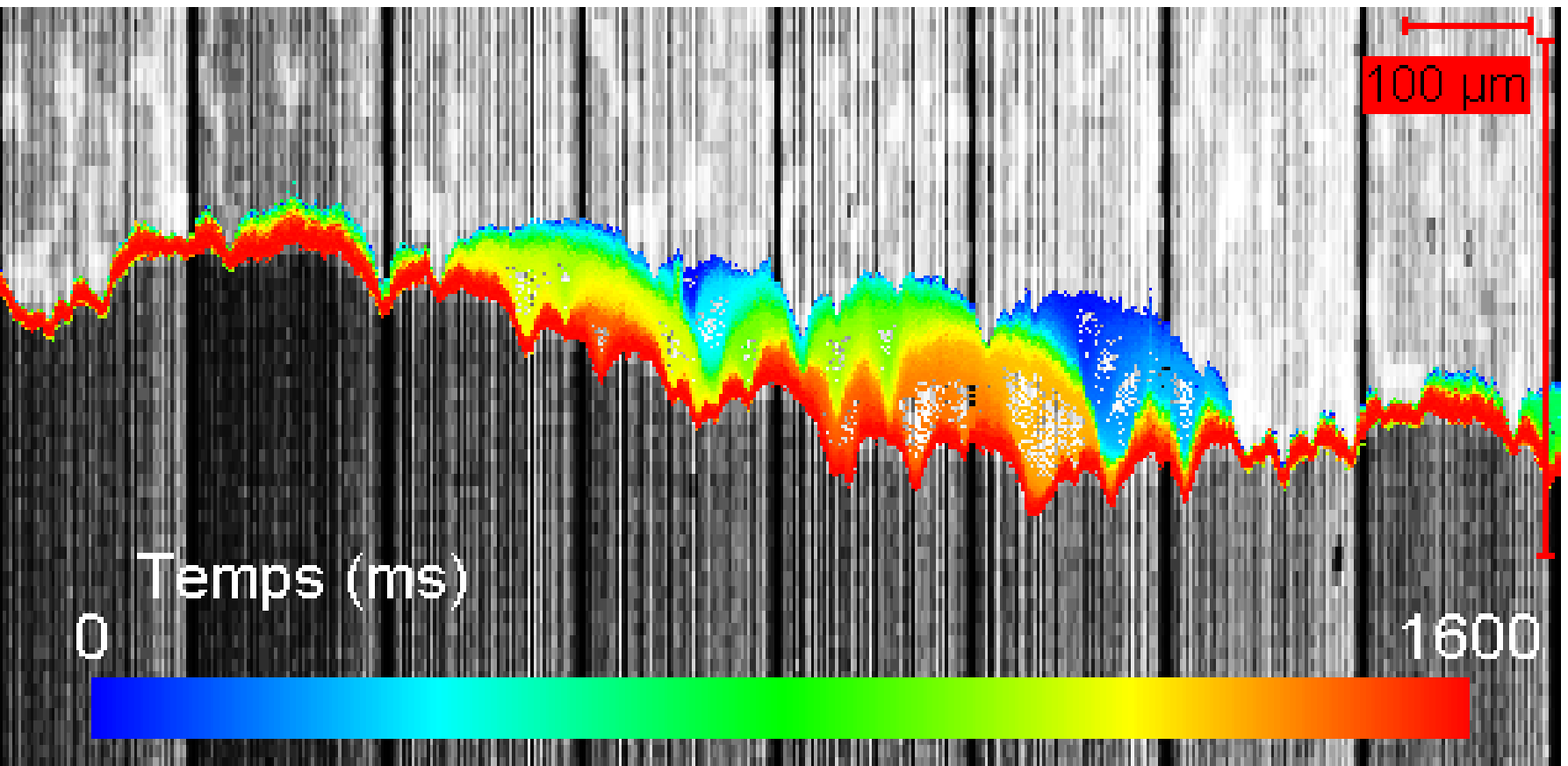}
\end{minipage}}}
\Caption{A contact line for the wetting of a disordered substrate by
Glycerine \protect\cite{MoulinetGuthmannRolley2002}. Experimental setup
(left). The disorder consists of randomly deposited islands of
Chromium, appearing as bright spots (top right). Temporal evolution of
the retreating contact-line (bottom right). Note the different scales
parallel and perpendicular to the contact-line. Pictures courtesy of
S.~Moulinet, with kind permission.}  \label{exp:contact-line}
\end{figure}
\begin{figure}[t]\label{f:vortex-lattic}
\centerline{\parbox{0.47\textwidth}{\fig{0.47\textwidth}{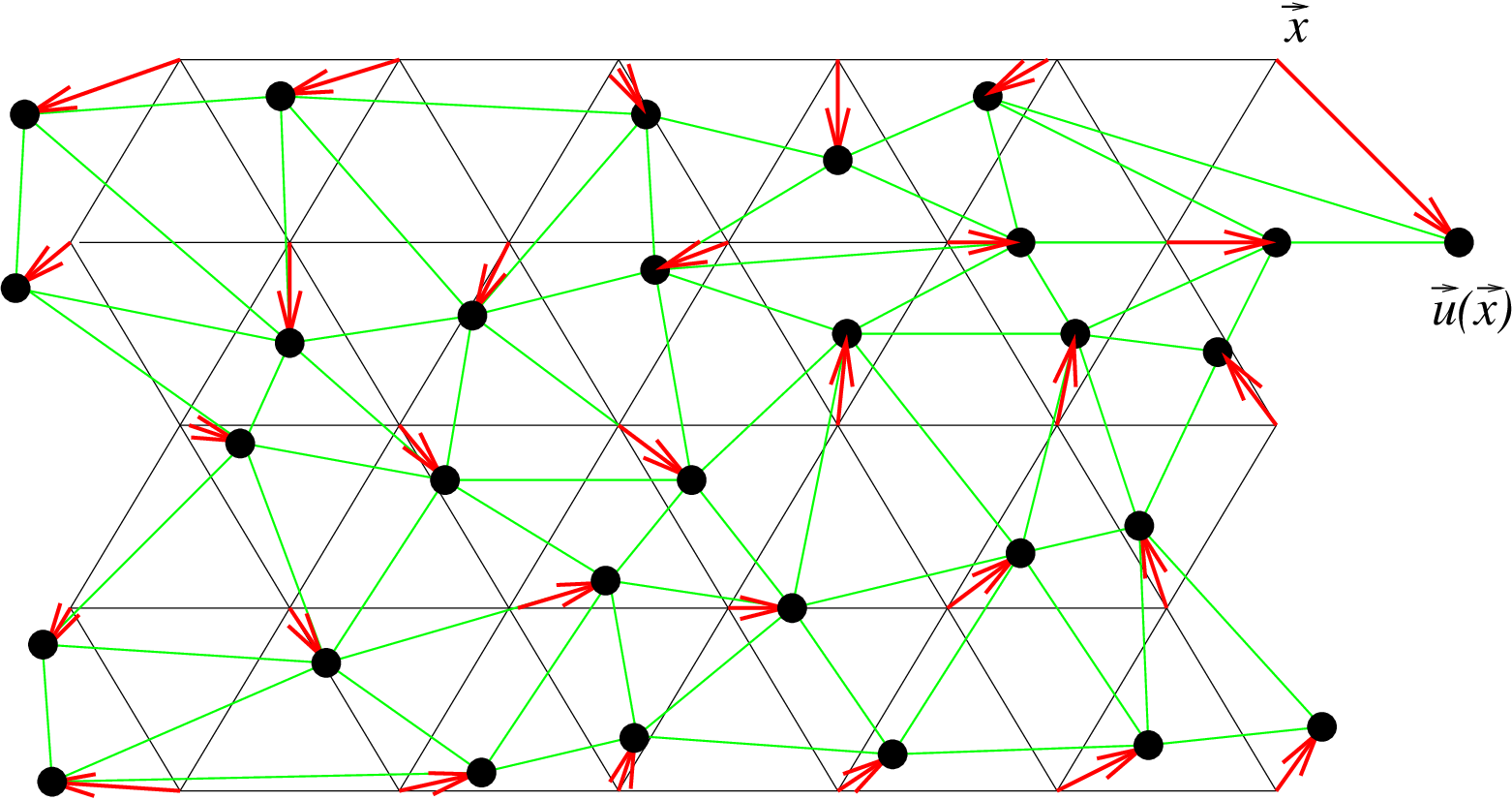}}}\smallskip
\Caption{Cartoon of an  elastic lattice (e.g.\ vortex lattice)
deformed by disorder. This is described by a vector $\vec u
(x)$.}
\end{figure}

Before developing the theory to treat elastic systems in a disordered
environment, let us give some physical realizations. The simplest one
is an Ising magnet. Imposing boundary conditions with all spins up at
the upper and all spins down at the lower boundary (see figure 1), at
low temperatures, a domain wall separates a region with spin up from a
region with spin down. In a pure system at temperature $T=0$, this
domain wall is completely flat.  Disorder can deform the domain wall,
making it eventually rough again. Two types of disorder are common:
random bond (which on a course-grained level also represents missing
spins) and random field (coupling of the spins to an external random
magnetic field). Figure 1 shows, how the domain wall is described by a
displacement field $u (x)$.  Another example is the contact line of
water (or liquid Helium), wetting a rough substrate, see figure
\ref{exp:contact-line}. (The elasticity is long range). A realization
with a 2-parameter displacement field $\vec{u} (\vec x) $ is the
deformation of a vortex lattice: the position of each vortex is
deformed from $\vec x$ to $\vec x+ \vec u (\vec x)$.  A 3-dimensional
example are charge density waves.

All these models have in common, that they are described
by a displacement field
\begin{equation}\label{u}
x\in \R^d \ \longrightarrow\  \vec u (x) \in \R^N
\ .
\end{equation}
For simplicity, we set $N=1$ in the following.  After some initial
coarse-graining, the energy ${\cal H}={\cal H}_{\mathrm{el}}+{\cal
H}_{\mathrm{DO}}$ consists out of two parts: the elastic energy
\begin{equation}
{\cal H}_{\mathrm{el}}[u] = \int \rmd ^d x \, \half \left( \nabla u
(x)\right)^2
\end{equation}
and the disorder
\begin{equation}\label{HDO}
{\cal H}_{\mathrm{DO}}[u] = \int \rmd ^{d} x \, V (x,u (x))\ .
\end{equation}
In order to proceed, we need to specify the  correlations of
disorder. Suppose that fluctuations $u$ in
the transversal direction scale  as
\begin{equation}\label{roughness}
\overline{\left[u (x)-u (y) \right]^{2}}  \sim  |x-y|^{2\zeta }
\end{equation}
with a roughness-exponent $\zeta <1$. Starting from a disorder
correlator
\begin{equation}
\overline{V (x,u)V (x',u')} = f (x-x') R (u-u')
\end{equation}
and performing one step in the RG-procedure, one has to rescale more
in the $x$-direction than in the $u$-direction. This will eventually
reduce $f (x-x')$ to a $\delta $-distribution, whereas the structure
of $R (u-u')$ remains visible.  We therefore choose as our
starting model
\begin{equation}\label{DOcorrelR}
\overline{V (x,u)V (x',u')} := \delta ^{d } (x-x') R (u-u')
\ .
\end{equation}
There are a couple of useful observables. We already mentioned the
roughness-exponent $\zeta $. The second is the renormalized
(effective) disorder. It will turn out that we actually have to keep
the whole disorder distribution function $R (u)$, in contrast to
keeping a few moments.  Other observables are higher correlation
functions or the free energy.

\section{Treatment of disorder}\label{treat disorder} Having defined
our model, we can now turn to the treatment of disorder. The problem
is to average not the partition-function, but the free energy over
disorder: $\overline{{\cal F}}=- k_{\mathrm{B}}T \, \overline{\ln Z} $. This can be
achieved by the beautiful {\em replica-trick}. The idea is to write
\begin{equation}
\ln {\cal Z} = \lim_{n\to 0} \frac{1}{n}\left( \rme^{n \ln {\cal Z}}-1
\right) = \lim_{n\to 0} \frac{1}{n}\left({\cal Z}^{n}-1 \right)
\end{equation}
and to interpret ${\cal Z}^{n}$ as the partition-function of an $n$
times replicated system. Averaging $\rme ^{-\sum _{a=1}^{n}{\cal
H}_{a}}$ over disorder then leads to the {\em replica-Hamiltonian}
\begin{equation}\label{H}
{\cal H}[u] = \frac{1}{T} \sum _{a=1}^{n}\int \rmd ^{d }x\, \half
\left(\nabla u_{a} (x) \right)^{2} -\frac{1}{2 T^{2}}  \sum
_{a,b=1}^{n} \int \rmd ^{d }x\, R (u_{a} (x)-u_{b} (x))\ .
\end{equation}
Let us stress that one could equivalently pursue a dynamic (see section \ref{s:dynamics}) or a
supersymmetric formulation (section \ref{a5}). We therefore should not, and in fact do
not encounter, problems associated with the use of the replica-trick,
as long as we work with a perturbative expansion in $R$.

\section{Flory estimates}\label{a1}
Four types of disorder have to be distinguished, resulting in different
universality classes:
\begin{itemize}
\item [ (i)] Random-Bond disorder (RB): short-range
correlated potential-potential correlations, i.e.\ short-range
correlated $R (u)$.
\item [ (ii)] Random-Field disorder (RF): short-range
correlated force-force correlator $\Delta (u):= -R'' (u)$. As the name
says, this disorder is relevant for Random-field systems, where the
disorder potential is the sum over all magnetic fields in say the
spin-up phase.
\item [(iii)] Generic long-range correlated disorder: $R (u)\sim |u|^{-\gamma
}$.
\item [(iv)] Random-Periodic disorder (RP): Relevant when the disorder couples
to a phase, as e.g.\ in charge-density waves. $R (u)=R (u+1)$,
supposing that $u$ is periodic with period 1.
\end{itemize}

To get an idea how large the roughness $\zeta$ becomes in these
situations, one compares the contributions of elastic energy and
disorder, and demands that they scale in the same way. This estimate
has first been used by Flory for self-avoiding polymers, and is therefore
called the Flory estimate. Despite the fact that Flory estimates are
conceptually crude, they often yield a rather good
approximation. For RB this gives for an $N$-component field $u$:
$\int_{x} (\nabla u)^{2} \sim \int_{x} \sqrt{\overline{VV}}$, or $
L^{d-2} u^2 \sim L^{d} \sqrt{L^{-d}u^{-N}} $, i.e.\ $u \sim L ^{\zeta
}$ with
\begin{equation}\label{a2}
\zeta_{\mathrm{Flory}}^{\mathrm{RB}} = \frac{4-d}{4+N}\ .
\end{equation}
For RF it is $R''$ that is short-ranged, and we obtain
\begin{equation}\label{a3}
\zeta_{\mathrm{Flory}}^{\mathrm{RF}} = \frac{4-d}{2+N}\ .
\end{equation}
For LR
\begin{equation}\label{a4}
\zeta_{\mathrm{Flory}}^{\mathrm{LR}} = \frac{4-d}{4+\gamma }
\end{equation}
For RP, the amplitude of $u$ is fixed, and thus $\zeta_{\mathrm{RP}}=0$.

\section{Dimensional reduction}\label{dimred} There is a beautiful and rather
mind-boggling theorem relating disordered systems to pure systems
(i.e.\ without disorder), which applies to a large class of systems,
e.g.\ random field systems and elastic manifolds in disorder. It is
called dimensional reduction and reads as
follows\cite{EfetovLarkin1977}:

\noindent {\underline{Theorem:}} {\em A $d$-dimensional disordered
system at zero temperature is equivalent to all orders in perturbation
theory to a pure system in $d-2$ dimensions at finite temperature. }
Moreover the temperature is (up to a constant) nothing but the width
of the disorder distribution. A simple example is the 3-dimensional
random-field Ising model at zero temperature; according to the theorem
it should be equivalent to the pure 1-dimensional Ising-model at
finite temperature. But it has been shown rigorously, that the former
has an ordered phase, whereas we have all solved the latter and we
know that there is no such phase at finite temperature. So what went
wrong? Let us stress that there are no missing diagrams or any such
thing, but that the problem is more fundamental: As we will see later,
the proof makes assumptions, which are not satisfied.  Nevertheless,
the above theorem remains important since it has a devastating
consequence for all perturbative calculations in the disorder: However
clever a procedure we invent, as long as we do a perturbative
expansion, expanding the disorder in its moments, all our efforts are
futile: dimensional reduction tells us that we get a trivial and
unphysical result. Before we try to understand why this is so and how
to overcome it, let us give one more example. Dimensional reduction
allows to calculate the roughness-exponent $\zeta $ defined in
equation (\ref{roughness}).  We know (this can be inferred from
power-counting) that the width $u$ of a $d$-dimensional manifold at
finite temperature in the absence of disorder scales as $u\sim
x^{(2-d)/2}$. Making the dimensional shift implied by dimensional
reduction leads to
\begin{equation}\label{zetaDR}
\overline{\left[ u (x)-u (0) \right]^{2}} \sim x^{4-d} \equiv x^{2\zeta }
\quad \mbox{i.e.}\quad \zeta =\frac{4-d}{2}\ .
\end{equation}

\section{The Larkin-length, and the role of temperature}\label{Larkin}
To understand the failure of dimensional reduction, let us turn to an
interesting argument given by Larkin \cite{Larkin1970}. He considers a
piece of an elastic manifold of size $L$. If the disorder has
correlation length $r$, and characteristic potential energy $\bar f$,
this piece will typically see a potential energy of strength
\begin{equation}
E_{\mathrm{DO}} = \bar f \left(\frac{L}{r} \right)^{\!\frac{d}{2}}\ .
\end{equation}
On the other hand, there is an elastic energy, which scales like
\begin{equation}
E_{\mathrm{el}} = c\, L^{d-2}\ .
\end{equation}
These energies are balanced at the  {\em Larkin-length} $L=L_{c}$
with
\begin{equation}
L_{c} = \left(\frac{c^{2}}{\bar f^{2}}r^{d} \right)^{\frac{1}{4-d}}
\ .
\end{equation}
More important than this value is the observation that in all
physically interesting dimensions $d<4$, and at scales $L>L_{c}$, the
membrane is pinned by disorder; whereas on small scales the elastic energy
dominates. Since the disorder has a lot of minima which are far apart
in configurational space but close in energy (metastability), the
manifold can be in either of these minimas, and the ground-state is no
longer unique. However exactly this is assumed in e.g.\ the proof of
dimensional reduction; as is most easily seen in its supersymmetric
formulation, see \cite{ParisiSourlas1979} and section \ref{a5}.

Another important question is, what the role of temperature is. In
(\ref{roughness}), we had supposed that $u$ scales with the systems
size, $u\sim L^{\zeta}$. From the first
term in (\ref{H}) we conclude that
\begin{equation}\label{a8}
T\sim L^{\theta}\ ,\qquad  \theta =d-2+2 \zeta
\end{equation}
Temperature is irrelevant when $\theta >0$, which is the
case for $d>2$, and when $\zeta >0$ even below. The RG fixed point we
are looking for will thus always be at zero temperature.

From the second term in (\ref{H}) we conclude that  disorder scales
as
\begin{equation}\label{R-scaling}
R\sim L^{d-4+4\zeta}\ .
\end{equation}
This is another way to see that $d=4$ is the upper critical
dimension.

\section{The functional renormalization group (FRG)}\label{FRG}

Let us now discuss a way out of the dilemma: Larkin's argument
(section \ref{Larkin}) or Eq.~(\ref{R-scaling}) suggests that $4$ is the upper critical
dimension. So we would like to make an $\epsilon =4-d$ expansion. On
the other hand, dimensional reduction tells us that the roughness is
$\zeta =\frac{4-d}{2}$ (see (\ref{zetaDR})). Even though this is
systematically wrong below four dimensions, it tells us correctly that
at the critical dimension $d=4$, where disorder is marginally
relevant, the field $u$ is dimensionless. This means that having
identified any relevant or marginal perturbation (as the disorder), we
find immediately another such perturbation by adding more powers of
the field. We can thus not restrict ourselves to keeping solely the
first moments of the disorder, but have to keep the whole
disorder-distribution function $R (u)$. Thus we need a {\em functional
renormalization group} treatment (FRG). Functional renormalization is
an old idea going back to the seventies, and can e.g.\ be found in
\cite{WegnerHoughton1973}.  For disordered systems, it was first
proposed in 1986 by D.\ Fisher \cite{DSFisher1986}.  Performing an
infinitesimal renormalization, i.e.\ integrating over a momentum shell
\`a la Wilson, leads to the flow $\partial _{\ell} R (u)$, with
($\epsilon =4-d$)
\begin{equation}\label{1loopRG}
\partial _{\ell} R (u) = \left(\epsilon -4 \zeta  \right) R (u) +
\zeta u R' (u) + \frac{1}{2} R'' (u)^{2}-R'' (u)R'' (0)\ .
\end{equation}
The first two terms come from the rescaling of $R$ in Eq.\
(\ref{R-scaling}) and $u$  respectively. The last two terms are the
result of the 1-loop calculations, which are derived in appendix
\ref{app:deriveRG}.

\begin{figure}[t]
\centerline{\fig{13.4cm}{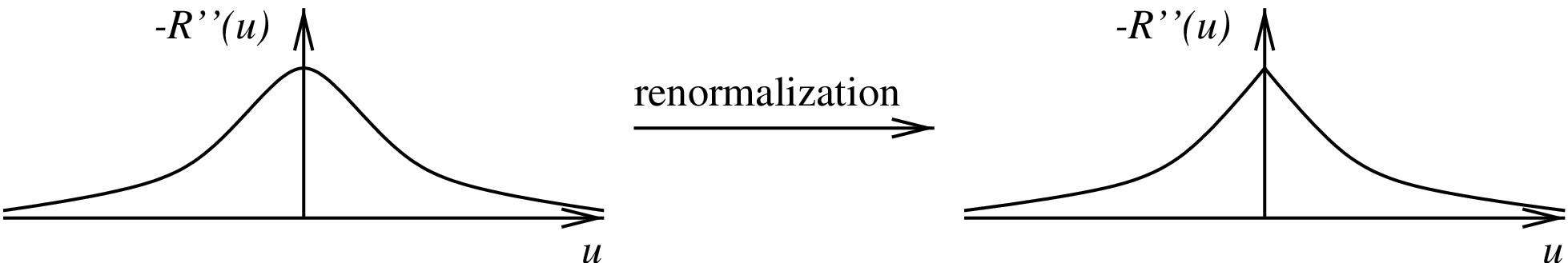}}
\Caption{Change of $-R'' (u)$ under renormalization and formation of
the cusp.} \label{fig:cusp}
\end{figure}

More important than the form of this equation is it actual solution,
sketched in figure \ref{fig:cusp}.
After some finite renormalization, the second derivative of the
disorder $R'' (u)$ acquires a cusp at $u=0$; the length at which this
happens is the Larkin-length. How does this overcome dimensional
reduction?  To understand this, it is interesting to study the flow of
the second and forth moment. Taking derivatives of (\ref{1loopRG})
w.r.t.\ $u$ and setting $u$ to 0, we obtain
\begin{eqnarray}
\partial_{\ell} R'' (0) &=& \left(\epsilon -2 \zeta  \right) R'' (0) +
R''' (0)^{2} \ \longrightarrow \ \left(\epsilon -2 \zeta  \right) R''
(0)\label{R2of0}\\
\partial_{\ell} R'''' (0) &=& \epsilon  R'''' (0) + 3 R'''' (0)^{2} +4 R'''
(0)R''''' (0)  \ \longrightarrow\ \epsilon  R'''' (0) + 3 R''''
(0)^{2}\label{R4of0}
\ .
\end{eqnarray}
Since $R (u)$ is an even function, and moreover the microscopic
disorder is smooth (after some initial averaging, if necessary), $R'''
(0)$ and $R''''' (0)$ are 0, which we have already indicated in Eqs.\
(\ref{R2of0}) and (\ref{R4of0}) . The above equations for $R'' (0)$
and $R'''' (0)$ are in fact closed.  Equation (\ref{R2of0}) tells us
that the flow of $R'' (0)$ is trivial and that $\zeta =\epsilon
/2\equiv \frac{4-d}{2}$. This is exactly the result predicted by
dimensional reduction. The appearance of the cusp can be inferred from
equation (\ref{R4of0}). Its solution is
\begin{equation}
R'''' (0)\ts _{\ell}= \frac{c\,\rme^ {\epsilon \ell }}{1-3\, c \left(\rme^
{\epsilon \ell} -1 \right)/ \epsilon }\ , \qquad c:= R'''' (0)\ts _{\ell=0}
\end{equation}
Thus after a finite renormalization $R'''' (0)$ becomes infinite: The
cusp appears. By analyzing the solution of the flow-equation
(\ref{1loopRG}), one also finds that beyond the Larkin-length $R''
(0)$ is no longer given by (\ref{R2of0}) with $R''' (0)^{2}=0$.  The
correct interpretation of (\ref{R2of0}), which remains valid after the
cusp-formation, is (for details see below)
\begin{equation}
\partial_{\ell} R'' (0) = \left(\epsilon -2 \zeta  \right) R'' (0)  +R''' (0^{+})^{2} \label{R2of0after}\ .
\end{equation}
Renormalization of the whole function thus overcomes dimensional
reduction.  The appearance of the cusp also explains why dimensional
reduction breaks down. The simplest way to see this is by redoing the
proof for elastic manifolds in disorder, which in the absence of
disorder is a simple Gaussian theory. Terms contributing to the
2-point function involve $R'' (0)$, $TR'''' (0)$ and higher
derivatives of $R (u)$ at $u=0$, which all come with higher powers of
$T$. To obtain the limit of $T\to 0$, one sets $T=0$, and only $R''
(0)$ remains. This is the dimensional reduction result. However we
just saw that $R'''' (0)$ becomes infinite. Thus $R'''' (0) T$ may
also contribute, and the proof fails.

\section{Measuring the cusp}\label{measurecusp}

Until now the function $R(u)$, a quantity central to the FRG, was
loosely described as an effective disorder correlator, which evolves
under coarse-graining towards a non-analytic shape. It turns out that
it can be given a precise definition as an {\em observable} \cite{LeDoussal2006b}. 
Hence it can {\em directly} be computed in the numerics, as we will discuss below,
and in principle, be measured in experiments. The cusp therefore is
not a theoretical artefact, but a real property of the system, related
to singularities or shocks, which arise in the landscape of pinning
forces. Moreover, these singularities are unavoidable for a glass with multiple
metastable states.

Consider our interface in a random potential, and add an external
quadratic potential well, centered around $w$:
\begin{eqnarray}
{\cal H}_{\mathrm{tot}}^{w}[u] = \frac{m^2}{2} (u(x)-w)^2 + {\cal
H}_{\mathrm{el}}[u] + {\cal H}_{\mathrm{DO}}[u]\ .
\end{eqnarray}
In each sample (i.e.\ disorder configuration), and given $w$, one
finds the minimum energy configuration. This 
ground state energy is
\begin{eqnarray}
\hat V(w) := \min_{u(x)} {\cal H}_{\mathrm{tot}}^{w}[u]\ .
\end{eqnarray}
It varies with $w$ as well as from sample to sample. Its second
cumulant
\begin{eqnarray}
\overline{ \hat V(w)  \hat V(w') }^c = L^d R(w-w') \label{defR}
\end{eqnarray}
defines a function $R(w)$ which is proven \cite{LeDoussal2006b} to
be the same function computed in the field theory, defined from the zero-momentum
action \cite{LeDoussalWieseChauve2003}.

Physically, the role of the well is to forbid the interface to wander
off to infinity. The limit of small $m$ is then taken to reach the
universal limit. The factor of volume $L^d$ is necessary, since the width
$\overline{u^2}$ of the interface in the well cannot grow much more
than $m^{-\zeta}$. This means that the interface is made of roughly
$L/L_m$ pieces of internal size $L_m \approx m$ pinned independently:
(\ref{defR}) hence expresses the central limit theorem and $R(w)$
measures the second cumulant of the disorder seen by any one of the
independent pieces.

\begin{figure}[t]\setlength{\unitlength}{1.4mm}
\fboxsep0mm \centerline{\mbox{\fig{10cm}{compareRFRBchaos}}}
\Caption{Filled symbols show numerical results for $Y(z)$, a
normalized form of the interface displacement correlator $-R''(u)$
[Eq.\ (\ref{defDe})], for $D=2+1$ random field (RF) and $D=3+1$ random
bond (RB) disorders. These suggest a linear cusp. The inset plots the
numerical derivative $Y'(z)$, with intercept $Y'(0)\approx -0.807$
from a quadratic fit (dashed line).  Open symbols plot the
cross-correlator ratio $Y_s(z)=\Delta_{12}(z)/\Delta_{11}(0)$ between
two related copies of RF disorder. It does not exhibit a cusp.  The
points are for confining wells with width given by $M^2=0.02$.
Comparisons to 1-loop FRG predictions (curves) are made with no
adjustable parameters. Reprinted from \cite{MiddletonLeDoussalWiese2006}.}  
\label{f:Alan1}
\end{figure}
The nice thing about (\ref{defR}) is that it can be measured. One
varies $w$ and computes (numerically) the new ground-state energy;
finallying averaging over many realizations. This has been performed
recently in \cite{MiddletonLeDoussalWiese2006} using a powerful
exact-minimization algorithm, which finds the ground state in a time
polynomial in the system size. In fact, what was measured there are
the fluctuations of the center of mass of the interface $u(w)=L^{-d}
\int \rmd^d x\, u_0(x;w)$:
\begin{eqnarray}
\overline{[w-u(w)] [w'-u(w')] }^c = m^{-4} L^{-d} \Delta(w-w')
\label{defDe}
\end{eqnarray}
\begin{figure}[b]
\centerline{\rotatebox{90}{\qquad\qquad ~~ $w-u_{w}$}
\includegraphics[width=8cm,viewport=125 255 450
440,clip]{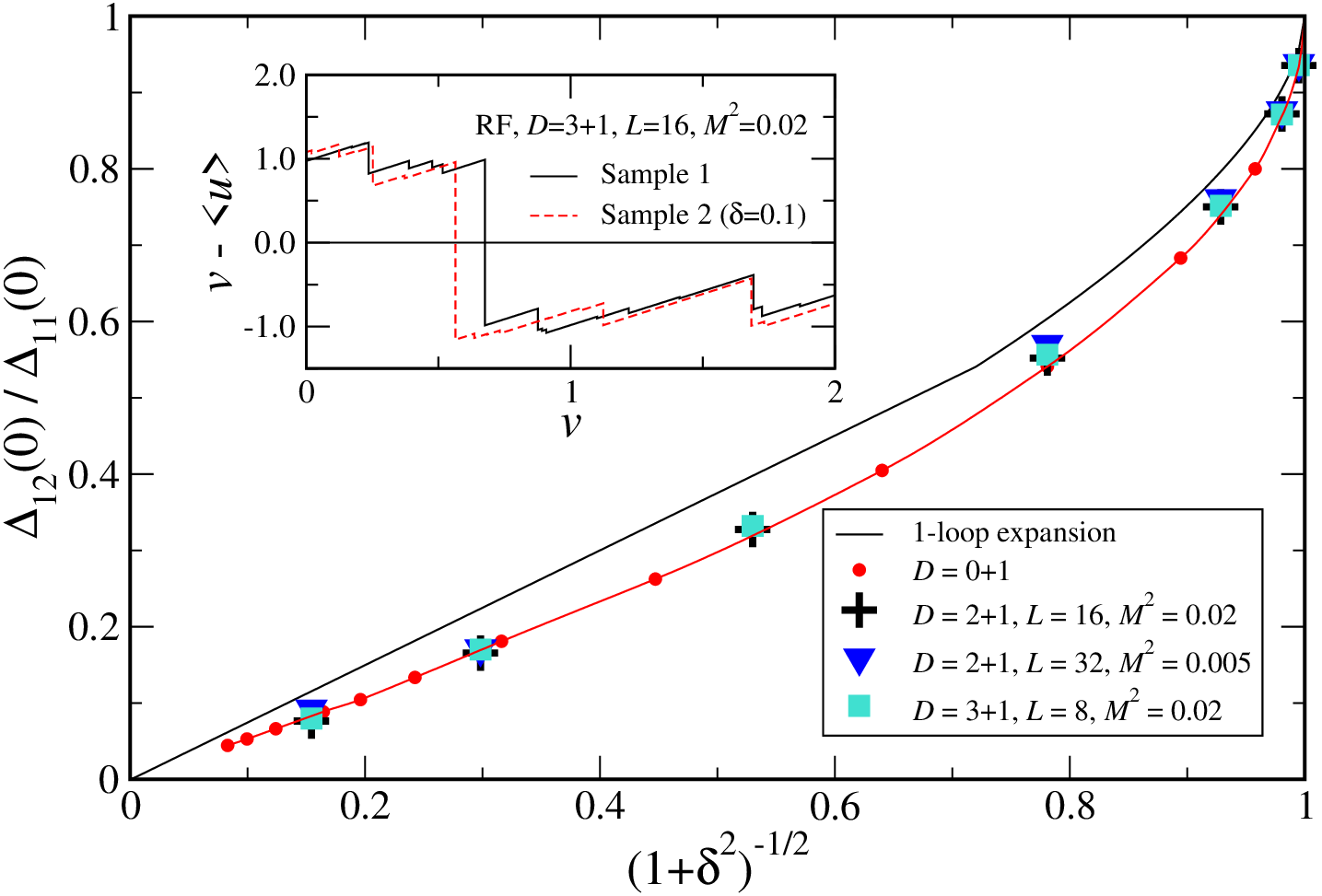}}
\centerline{\qquad $w$}
\Caption{Discontinuous positions, ``shocks'', in $w-u_{w}$ as a function
of $w$. Reprinted from \cite{MiddletonLeDoussalWiese2006}.}
\label{f:Alan-Shocks}
\end{figure}%
which measures directly the correlator of the pinning force
$\Delta(u)=-R''(u)$. To see why it is the total force, write the
equilibrium condition for the center of mass $m^2 [w-u(w)] + L^{-d}
\int \rmd^d x\, F(x,u)=0$ (the elastic term vanishes if we use periodic
b.c.). The result is represented in figure \ref{f:Alan1}. It is most
convenient to plot the function $Y=\Delta(u)/\Delta(0)$ and normalize
the $u$-axis to eliminate all non-universal scales.
The plot in figure \ref{f:Alan1} is free of any parameter. It has
several remarkable features. First, it clearly shows that a linear
cusp exists in any dimension. Next it is very close to the 1-loop
prediction. Even more remarkably, as detailed in
\cite{MiddletonLeDoussalWiese2006}, the statistics is good enough to
reliably compare the deviations to the 2-loop predictions obtained
in section \ref{2loop}.

What is the physics of the cusp in $\Delta(u)$? One easily sees in a
zero-dimensional model, i.e.\ a particle on a line, 
$d=0$, that as $w$ increases, the position of the minimum $u(w)$
increases smoothly, except at some discrete set of positions $w=w_s$,
where the system switches abruptly between two distant
minima. Formally, one can show that the landscape of the force $-\hat
V'(w)$ evolves, as the mass is lowered, according to a Burgers
equation, known to develop finite-time singularities called
``shocks''. For details on this mapping see
\cite{BalentsBouchaudMezard1996,LeDoussal2006b}. For an interface
these shocks also exist, as can be seen on figure
\ref{f:Alan-Shocks}.  Note that when we vary the position $w$ of the
center of the well, it is not a real motion. It just means to find the
new ground state for each $w$.  Literally ``moving'' $w$ is another
very interesting possibility, and will be discussed in section
\ref{s:dynamics} devoted to depinning
\cite{LeDoussalWiese2006a,RossoLeDoussalWiese2006a}.

\section{Rounding the cusp}\label{s:Rounding the cusp} As we have
seen, a cusp non-analyticity necessary arises at zero temperature, due
to the switch-over between many metastable states.  Interestingly,
this cusp can be rounded by several effects: By non-zero temperature
$T>0$, chaos, or a non-zero driving velocity (in the dynamics
discussed below). It is easy to include the effect of temperature in
the FRG equation to one loop \cite{ChauveGiamarchiLeDoussal2000}:
\begin{equation}\label{tempRG}
\partial _{\ell} R (u) = \left(\epsilon -4 \zeta  \right) R (u) +
\zeta u R' (u) + \frac{1}{2} R'' (u)^{2}-R'' (u)R'' (0) + \tilde T_\ell
R''(u) \ .
\end{equation}
$\tilde T_\ell = T \rme^{- \theta \ell}$ is the dimensionless
temperature. It finally flows to zero, since temperature is an irrelevant
variable as discussed above. Although irrelevant, it has some profound
effect. Clearly the temperature in (\ref{tempRG}) acts as a diffusive
term smoothening the cusp. In fact, at non-zero temperature there
never is a cusp, and $R(u)$ remains analytic. The convergence to the
fixed point is non-uniform. For $u$
fixed, $R(u)$ converges to the zero-temperature fixed point, except
near $u=0$, or more precisely in a boundary layer of size $u \sim
\tilde T_\ell$, which shrinks to zero in the large-scale limit.
Non-trivial consequences are: The curvature blows up as $R''''(0) \sim
\rme^{\theta \ell}/T \sim L^\theta/T$. One can show that this is related
to the existence of thermal excitations (``droplets'') in the statics
\cite{BalentsLeDoussal2004} and of ``barriers'' in the dynamics, which
grow as $L^\theta$ \cite{BalentsLeDoussal2003}.

Another case where rounding occurs is for ``disorder chaos''. Disorder
chaos is the possibility of a system to have a completely different
ground state at large scales, upon a very slight change in the
microscopic disorder (for instance changing slightly the magnetic
field in a superconductor). Not all types of disorder exhibit
chaos. Its presence in spin glasses is still debated. Recently it was
investigated for elastic manifolds, using FRG
\cite{LeDoussal2006a}. One studies a model with two copies, $i=1,2$,
each seeing slightly different disorder energies $V_{i} (x,u (x))$ in
Eq.~(\ref{HDO}).  The latter are mutually correlated gaussian random
potentials with a correlation matrix
\begin{eqnarray}\label{ViVj}
 \overline{V_i(x,u) V_j(x',u')} =  \delta^d(x-x') R_{ij}(u-u')\ .
\end{eqnarray}
At zero temperature, the FRG equations for $R_{11} (u) =R_{22} (u)$
are the same as in (\ref{1loopRG}). The one for the cross-correlator
$R_{12}(u)$ satisfies the same equation as (\ref{tempRG}) above, with
$\tilde T_\ell$ is replaced by $\hat T:=R''_{12}(0)-R''_{11}(0)$. It is
some kind of fictitious temperature, whose flow must be determined
self-consistently from the two FRG equations. As in the case of a real
temperature, it results in a rounding of the cusp. The physics of that
is apparent from figure \ref{f:Alan-Shocks}, which shows the set of
shocks in two correlated samples. Since they are slightly and randomly
displaced from each other, the cusp is rounded.

Chaos is obtained when $\hat T$ grows with scale, and occurs on scales
larger than the so-called overlap length. The mutual correlations
$C_{ij}(x-x') = \overline{\left< [u^i(x) - u^i(x')] [u^j(x) -
u^j(x')]\right>}$ behave as $C_{ij}(x) = x^{2 \zeta} f(\delta
x^\alpha)$, where $\delta$ quantifies the difference between the two
disorders at the microscopic level. $C_{ij} (x)$ decays at large
distance as $C_{ij}(x) \sim x^{2 \zeta - \mu}$ \cite{LeDoussal2006a}.

\section{Beyond 1 loop}\label{beyond1loop} Functional renormalization
has successfully been applied to a bunch of problems at 1-loop
order. From a field theory, we however demand more. Namely that
it\medskip

$\bullet$ allows for systematic corrections beyond 1-loop order\smallskip

$\bullet$ be renormalizable\smallskip

$\bullet$ and thus allows to make universal predictions.\medskip

\noindent However, this has been a puzzle since 1986, and it has even
been suggested that the theory is not renormalizable due to the
appearance of terms of order $\epsilon ^{\frac{3}{2}}$
\cite{BalentsDSFisher1993}. Why is the next order so complicated? The
reason is that it involves terms proportional to $R''' (0)$. A look at
figure 3 explains the puzzle. Shall we use the symmetry of $R (u)$ to
conclude that $R''' (0)$ is 0? Or shall we take the left-hand or
right-hand derivatives, related by
\begin{equation}
R''' (0^{+}) := \lim_{{u>0}\atop {u\to 0}} R ''' (u) = -
\lim_{{u<0}\atop {u\to 0}} R ''' (u) =:- R''' (0^{-}) .
\end{equation}
In the following, we will present our solution of this puzzle, obtained
at 2-loop order,  at large
$N$, and in the driven dynamics.

\section{Results at 2-loop order}\label{2loop} For the flow-equation
at 2-loop order, the result is
\cite{LeDoussalWieseChauve2003,ChauveLeDoussalWiese2000a,Scheidl2loopPrivate,DincerDiplom,ChauveLeDoussal2001}
\begin{eqnarray}\label{2loopRG}
\partial _{\ell} R (u) &=& \left(\epsilon -4 \zeta  \right) R (u) +
\zeta u R' (u) + \frac{1}{2} R'' (u)^{2}-R'' (u)R'' (0) \nn \\
&& + \frac{1}{2}\left(R'' (u)-R'' (0) \right)R'''
(u)^{2}-\frac{1}{2}R''' (0^{+})^{2 } R'' (u) \ .
\end{eqnarray}
The first line is the result at 1-loop order, already given in
(\ref{1loopRG}). The second line is new. The most interesting term is
the last one, which involves $R''' (0^{+})^{2}$ and which we therefore
call {\em anomalous}.  The hard task is to fix the prefactor
$(-\frac{1}{2})$.   We have found five different prescriptions to
calculate it: The sloop-algorithm, recursive construction,
reparametrization invariance, renormalizability, and potentiality
\cite{ChauveLeDoussalWiese2000a,LeDoussalWieseChauve2003}. For lack
of space, we restrain our discussion to the last two ones. At 2-loop
order the following diagram appears
\begin{equation}\label{rebi}
\diagram{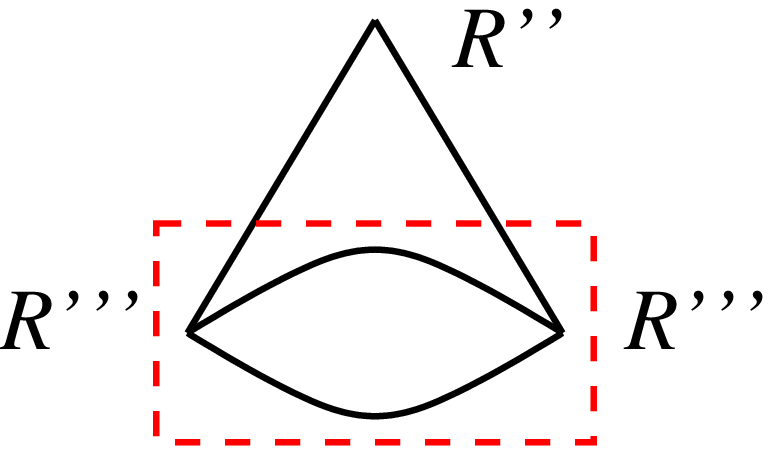}\  \longrightarrow\ \frac{1}{2}\left(R'' (u)-R'' (0)
\right)R'''
(u)^{2} -\half R'' (u)R''' (0^{+})^{2}
\end{equation}
leading to the anomalous term. The integral (not written here)
contains a sub-divergence, which is indicated by the
box. Renormalizability demands that its leading divergence (which is
of order $1/\epsilon ^{2}$) be canceled by a 1-loop counter-term. The
latter is unique thus fixing the prefactor of the anomalous term. (The
idea is to take the 1-loop correction $\delta R$ in Eq.~(\ref{80}) and
replace one of the $R''$ in it by $\delta R''$ itself, which the
reader can check to leading to the terms given in (\ref{rebi}) plus
terms which only involve even derivatives.)

Another very physical demand is that the problem remain potential,
i.e.\ that forces still derive from a potential. The force-force
correlation function being $-R'' (u)$, this means that the flow of
$R' (0)$ has to be strictly 0. (The simplest way to see this is to
study a periodic potential.) From (\ref{2loop}) one can check that
this does not remain true if one changes the prefactor of the last
term in (\ref{2loop}); thus fixing it.

Let us give some results for cases of physical interest. First of all,
in the case of a periodic potential, which is relevant for
charge-density waves, the fixed-point function can be calculated
analytically as (we choose period 1, the following is for $u\in
\left[0,1 \right]$)
\begin{equation}
R^{*} (u) = - \left(\frac{\epsilon }{72}+\frac{\epsilon ^{2}}{108}+O
(\epsilon ^{3}) \right) u^{2} (1-u)^{2} +\mbox{const.}
\end{equation}
This leads to a universal amplitude.  In the case of random-field
disorder (short-ranged force-force correlation function) $\zeta
=\frac{\epsilon }{3}$, equivalent to the Flory estimate (\ref{a3}). 
For random-bond disorder (short-ranged
potential-potential correlation function) we have to solve
(\ref{2loopRG}) numerically, with the result 
\begin{equation}
\zeta = 0.208 298 04
\epsilon +0.006858 \epsilon ^{2} + O(\epsilon^{3})\ .
\end{equation} This compares well with numerical
simulations, see figure \ref{fig:numstat}. It is also surprisingly close, but distinctly different, from the Flory estimate (\ref{a2}), $\zeta=\epsilon/5$. 

\begin{figure}\centerline{\small
\begin{tabular}{|c|c|c|c|c|}
\hline
$\zeta _{\rm}$ & one loop & two loop & estimate &
simulation and exact\\
\hline
\hline
$d=3$  & 0.208 &  0.215  & $0.215\pm 0.01$  &
$0.22\pm 0.01$ \cite{Middleton1995}  \\
\hline
$d=2$ &0.417 &0.444 &$0.42\pm 0.02$ &  $0.41\pm 0.01$ \cite{Middleton1995} \\
\hline
$d=1$ & 0.625 & 0.687 &  $0.67\pm 0.02$ & $2/3$ \\
\hline
\end{tabular}}\medskip
\Caption{Results for $\zeta $ in the random bond case.}\label{fig:numstat}
\end{figure}

\section{Finite $N$}\label{s:finiteN} Up to now, we have studied the
functional RG for one component $N=1$. The general case of
finite $N$ is more difficult to handle, since derivatives of the
renormalized disorder now depend on the direction, in which this
derivative is taken. Define amplitude $u:=|\vec u|$ and direction
$\hat u:= \vec u/|\vec u|$ of the field. Then deriving the latter
variable leads to terms proportional to $1/u$, which are diverging in
the limit of $u\to 0$. This poses additional problems in the
calculation, and it is a priori not clear that the theory at $N\neq1$
exists, supposed this is the case for $N=1$. At 1-loop order
everything is well-defined \cite{BalentsDSFisher1993}. We have found a
consistent RG-equation at 2-loop order \cite{LeDoussalWiese2005a}:
\begin{figure}[b]
\centerline{{\unitlength1mm
\begin{picture} (90,55)
\put(0,0){\fig{85mm}{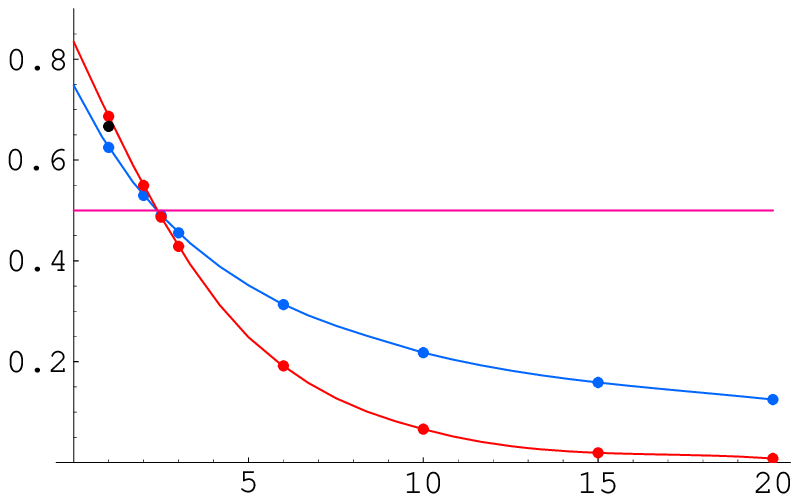}}
\put(5,52){$\zeta $}
\put(86,3){$N$}
\put(70,13){1-loop}
\put(30,7){2-loop}
\end{picture}}
}
\Caption{Results for the roughness $\zeta$ at 1- and 2-loop order, as
a function of the number of components $N$.}  \label{f:Ncomp}
\end{figure}
\begin{eqnarray}\label{2loopFPENcomp}
\partial_{\ell } R(u) &=& (\epsilon - 4 \zeta) R(u) + \zeta u R'(u)
+\frac{1}{2} R''(u)^2 - R''(0) R''(u) +\frac{N-1}{2} \frac{R'(u)}{u}
\left(\frac{R'(u)}{u} - 2 R''(0)\right)
\nn \\
&&+\frac{1}{2} \left( R''(u) - R''(0) \right) \,{R''' (u)}^2
+\frac{N{-}1}{2} \frac{{\left( R'(u) {-} uR''(u) \right) }^2\, ( 2
R'(u) {+} u(R''(u) {-}3 R''(0) ) )}{u^5}
\nonumber \\
&&  -R'''(0^{+})^{2} \left[\frac{N+3}{8}R''(u)+\frac{N-1}{4}\frac{R'(r)}{u} \right]
\ .
\end{eqnarray}
The first line is the 1-loop equation, given in
\cite{BalentsDSFisher1993}. The second and third line represent the
2-loop equation, with   the new anomalous terms proportional to $R'''
(0^{+})^{2}$ (third line).

The fixed-point equation (\ref{2loopFPENcomp}) can be integrated
numerically, order by order in $\epsilon$. The result, specialized to
directed polymers, i.e.\ $\epsilon =3$ is plotted on figure
\ref{f:Ncomp}.  We see that the 2-loop corrections are rather big at
large $N$, so some doubt on the applicability of the latter down to
$\epsilon=3$ is advised. However both 1- and 2-loop results reproduce
well the two known points on the curve: $\zeta =2/3$ for $N=1$ and
$\zeta =0$ for $N=\infty$. The latter result will be given in section
\ref{largeN}. Via the equivalence \cite{KPZ} of the directed-polymer
problem in $N$ dimensions treated here and the KPZ-equation of
non-linear surface growth in $N$ dimensions, which relate the
roughness exponent $\zeta$ of the directed polymer to the dynamic
exponent $z_{\mathrm{KPZ}}$ in the KPZ-equation via $\zeta
=\frac{1}{z_{{\mathrm{KPZ}}}}$, we know that $\zeta (N=1)=2/3$.

The line $\zeta =1/2$ given on figure \ref{f:Ncomp} plays a special
role: In the presence of thermal fluctuations, we expect the
roughness-exponent of the directed polymer to be bounded by $\zeta \ge
1/2$. In the KPZ-equation, this corresponds to a dynamic exponent
$z_{\mathrm{KPZ}}=2$, which via the exact scaling relation
$z_{\mathrm{KPZ}}+\zeta_{\mathrm{KPZ}}=2$ is an upper bound in the
strong-coupling phase. The above data thus strongly suggest that there
exists an upper critical dimension in the KPZ-problem, with
$d_{\mathrm{uc}}\approx 2.4$. Even though the latter value might be
an underestimation, it is hard to imagine what can go wrong {\em
qualitatively} with this scenario. The strongest objections will
probably arise from numerical simulations, such as
\cite{MarinariPagnaniParisi2000}. However the latter use a discrete
RSOS model, and the exponents are measured for interfaces, which in
large dimensions have the thickness of the discretization size,
suggesting that the data are far from the asymptotic regime. We thus
strongly encourage better numerical simulations on a continuous model,
in order to settle this issue.

\section{Large $N$}\label{largeN} In the last sections, we have
discussed renormalization in a loop expansion, i.e.\ expansion in
$\E$. In order to independently check consistency, it is good to have a
non-perturbative approach. This is achieved by the large-$N$ limit,
which can be solved analytically and to which we turn now. We start
from
\begin{eqnarray}\label{HlargeN}
{\cal H}[\vec u,\vec j ] &=& \frac{1}{2T} \sum _{a=1}^{n}\int_{x}
 \vec u_{a} (x)\left(-\nabla^{2}{+}m^{2} \right) \vec u_{a} (x) - \sum
_{a=1}^{n}\int_{x} \vec{j}_{a} (x)\vec{u}_{a} (x)  \nn \\
&&   -\frac{1}{2 T^{2}}  \sum
_{a,b=1}^{n} \int_x B \left((\vec u_{a} (x)-\vec u_{b} (x))^{2} \right)\ .
\end{eqnarray}
where in contrast to (\ref{H}), we use an $N$-component field $\vec{u}
$. For $N=1$, we identify $B (u^{2} )=R (u)$. We also have added a
mass $m$ to regularize the theory in the infra-red and a source
$\vec{j} $ to calculate the effective action $\Gamma (\vec u) $ via a
Legendre transform. For large $N$ the saddle-point equation reads
\cite{LeDoussalWiese2001}
\begin{equation}\label{saddlepointequation}
\tilde B' (u_{ab}^{2}) = B' \left(u_{ab}^{2}+2 T I_{1} + 4
I_{2} [\tilde B' (u_{ab}^{2})-\tilde B' (0)] \right)\ .
\end{equation}
This equation gives the derivative of the effective (renormalized)
disorder $\tilde B$ as a function of the (constant) background field
$u_{ab}^{2}= (u_{a}-u_{b})^{2}$ in terms of: the derivative of the
microscopic (bare) disorder $B$, the temperature $T$ and the integrals
$I_{n}:= \int_{k}\frac{1}{\left(k^{2}+m^{2} \right)^{n}}$.

The saddle-point equation can again be turned into a closed functional
renormalization group equation for $\tilde B$ by taking the derivative
w.r.t.\ $m$:
\begin{equation}\hspace{-0.9 cm}
\partial _{\ell}\tilde B (x)\equiv -\frac{m \partial }{\partial m}\tilde
B (x) =\left(\epsilon -4\zeta \right)\! \tilde B (x) + 2 \zeta x
\tilde B' (x)+\frac{1}{2}\tilde B' (x)^{2}-\tilde B' (x) \tilde B'
(0)+ \frac{\epsilon\, T \tilde B' (x)}{\epsilon +\tilde B'' (0)}\,\,\,
\end{equation}
This is a complicated nonlinear partial differential equation. It is
therefore surprising, that one can find an analytic solution. (The
trick is to write down the flow-equation for the inverse function of
$\tilde B' (x)$, which is linear.) Let us only give the results of
this analytic solution: First of all, for long-range correlated
disorder of the form $\tilde B' (x)\sim x^{-\gamma }$, the exponent
$\zeta $ can be calculated analytically as $\zeta =\frac{\epsilon }{2
(1+\gamma )}\ . $ It agrees with the replica-treatment in
\cite{MezardParisi1991},  the 1-loop treatment in
\cite{BalentsDSFisher1993}, and the Flory estimate (\ref{a4}). For short-range correlated disorder,
$\zeta =0$.  Second, it demonstrates that before the Larkin-length,
$\tilde B (x)$ is analytic and thus dimensional reduction
holds. Beyond the Larkin length, $\tilde B'' (0)=\infty $, a cusp
appears and dimensional reduction is incorrect. This shows again that
the cusp is not an artifact of the perturbative expansion, but an
important property even of the exact solution of the problem (here in
the limit of large $N$).

\section{Relation to Replica Symmetry Breaking (RSB)}\label{s:RSB} There
is another treatment of the limit of large $N$ given by M\'ezard and
Parisi \cite{MezardParisi1991}. They start from (\ref{HlargeN}) but
{\em without}\/ a source-term $j$. In the limit of large $N$, a
Gaussian variational ansatz of the form
\begin{eqnarray}\label{HlargeNMP}
{\cal H}_{\mathrm g}[\vec u] &=& \frac{1}{2T} \sum _{a=1}^{n}\int_{x}
 \vec u_{a} (x)\left(-\nabla^{2}{+}m^{2} \right) \vec u_{a} (x)
   -\frac{1}{2 T^{2}}  \sum
_{a,b=1}^{n} \sigma_{ab} \, \vec u_{a} (x)\vec u_{b} (x)
\end{eqnarray}
becomes exact. The art is to make an appropriate ansatz for
$\sigma_{ab}$. The simplest possibility, $\sigma _{ab}=\sigma $ for
all $a\neq b$ reproduces the dimensional reduction result, which
breaks down at the Larkin length. Beyond that scale, a replica
symmetry broken (RSB) ansatz for $\sigma _{ab}$ is suggestive. To this
aim, one can break $\sigma _{ab} $ into four blocks of equal size,
choose one (variationally optimized) value for the both outer diagonal
blocks, and then iterate the procedure on the diagonal blocks,
resulting in
\begin{equation}\label{RSB}
\sigma_{ab} =
\left(\,\parbox{.25\textwidth}{\fig{.25\textwidth}{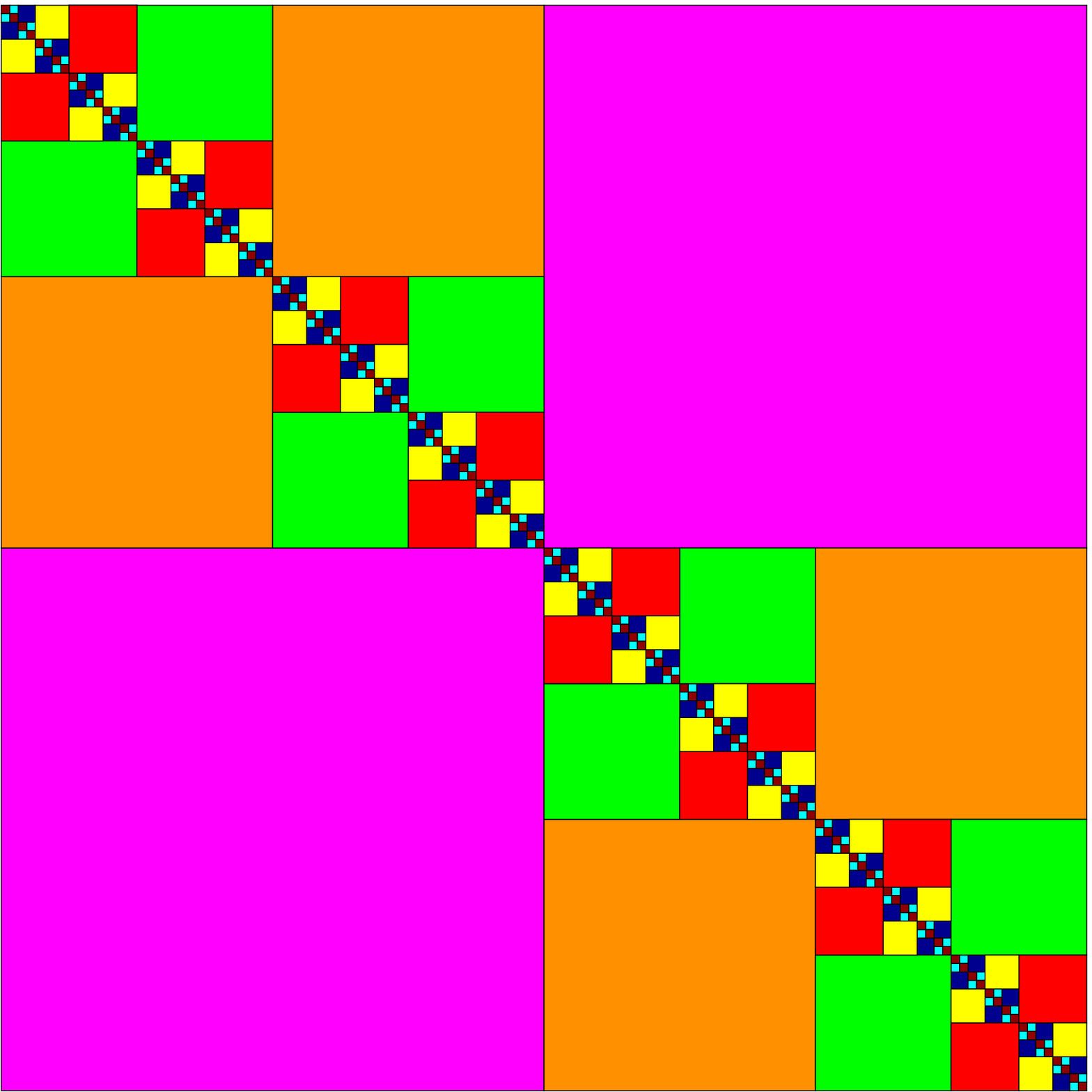}}\,\right)\ .
\end{equation}\begin{figure}[b]
\centerline{\fig{8cm}{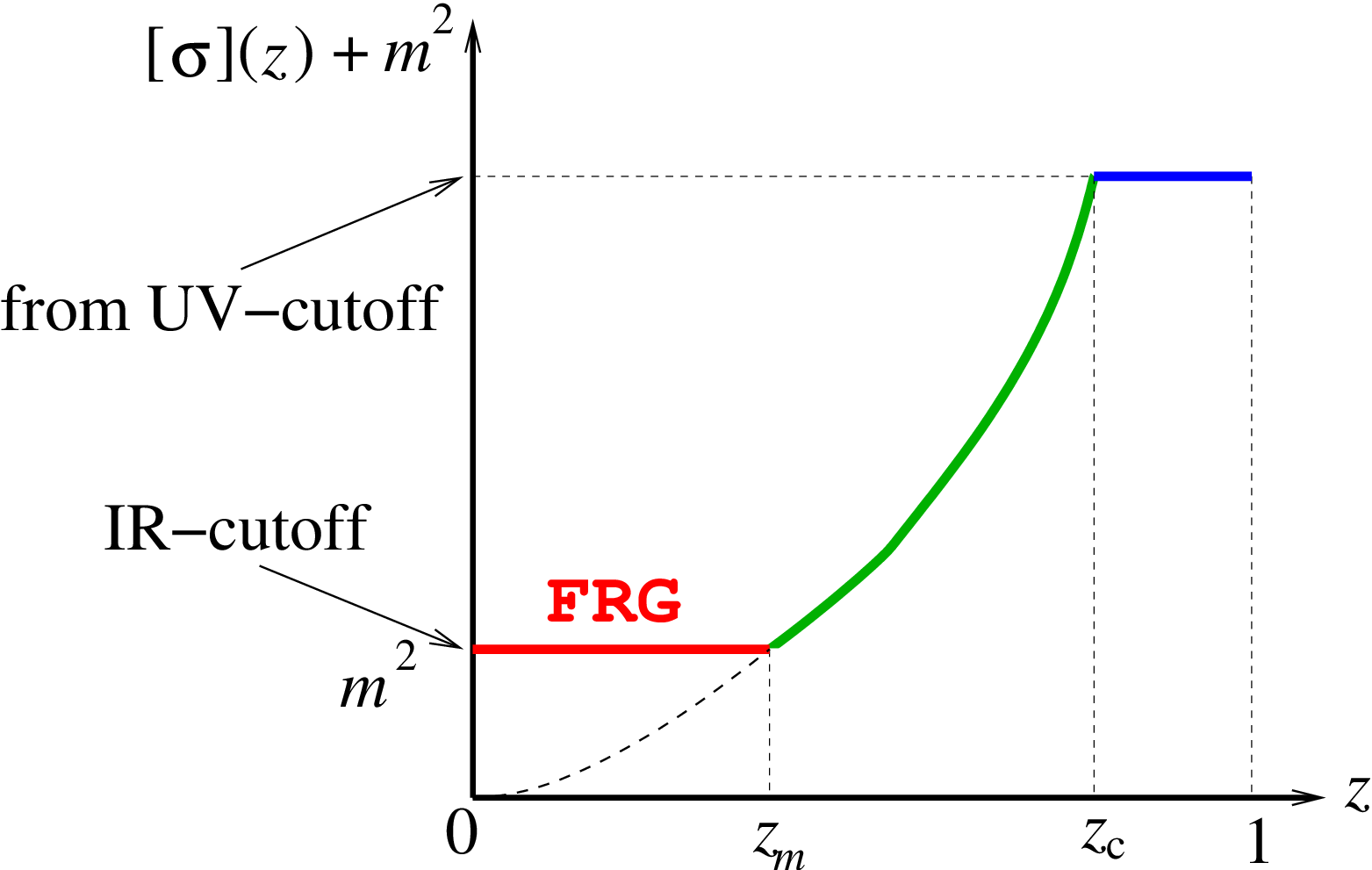}}
\Caption{The function $\left[\sigma \right] (u)+m^{2}$ as given in
\protect\cite{MezardParisi1991}.} \vspace{-0.1cm}\label{fig:MP-function}
\end{figure}%
One finds that the more often one iterates, the better the result becomes. In
fact, one has to repeat this procedure infinite many times. This seems
like a hopeless endeavor, but Parisi has shown that the infinitely
often replica symmetry broken matrix can be parameterized by a
function $[\sigma] (z)$ with $z\in \left[0,1 \right]$. In the
SK-model, $z$ has the interpretation of an overlap between
replicas. While there is no such simple interpretation for the model
(\ref{HlargeNMP}), we retain that $z=0$ describes distant states,
whereas $z=1$ describes nearby states. The solution of the large-$N$
saddle-point equations leads to the curve depicted in figure 6.
Knowing it, the 2-point function is given by
\begin{equation}\label{RSBformula}
\left< u_{k}u_{-k} \right>=\frac{1}{k^{2}+m^{2}}\left(1+\int_{0}^{1}
\frac{\rmd z}{z^{2}} \frac{\left[\sigma \right]
(z)}{k^{2}+\left[\sigma \right] (z)+m^{2}} \right)\ .
\end{equation}
The important question is: What is the relation between the two
approaches, which both declare to calculate the same 2-point function?
Comparing the analytical solutions, we find that the 2-point function
given by FRG is the same as that of RSB, if in the latter expression
we only take into account the contribution from the most distant
states, i.e.\ those for $z$ between 0 and $z_{m}$ (see figure
\ref{fig:MP-function}). To understand why this is so, we have to
remember that the two calculations were done under quite different
assumptions: In contrast to the RSB-calculation, the FRG-approach
calculated the partition function in presence of an external field
$j$, which was then used to give via a Legendre transformation the
effective action. Even if the field $j$ is finally tuned to 0, the
system will remember its preparation, as is the case for a magnet:
Preparing the system in presence of a magnetic field will result in a
magnetization which aligns with this field. The magnetization will
remain, even if finally the field is turned off. The same phenomena
happens here: By explicitly breaking the replica-symmetry through an
applied field, all replicas will settle in distant states, and the
close states from the Parisi-function $\left[\sigma \right] (z)+m^{2}$
(which describes {\em spontaneous} RSB) will not contribute.  However,
we found that the full RSB-result can be reconstructed by remarking
that the part of the curve between $z_{m}$ and $z_{c}$ is independent
of the infrared cutoff $m$, and then integrating over $m$
\cite{LeDoussalWiese2001} ($m_{c}$ is the mass corresponding to
$z_{c}$):
\begin{equation}\label{RSB=intFRG}
\left< u_{k}u_{-k} \right>\Big|^{\mathrm{RSB}}_{k=0} =\frac{\tilde
R'_{m}(0)}{m^{4}} +\int_{m}^{m_{c}} \frac{\rmd \tilde
R'_{\mu}(0)}{\mu^{4}} + \frac{1}{m_{c}^{2}}-\frac{1}{m^{2}}\ .
\end{equation}
We also note that a similar effective action  has been proposed in
\cite{BalentsBouchaudMezard1996}. While it agrees qualitatively, it
does not reproduce the correct FRG 2-point function, as
it should.

\section{Corrections at order $1/N$}\label{sec:1overN}
In a graphical notation, we find \cite{LeDoussalWiese2004a}
\begin{eqnarray}
\delta B^{(1)}&=&
\!\!\diagram{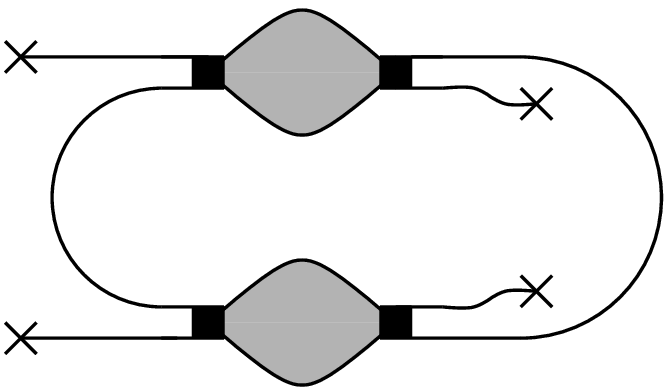}\!\!+\!\!\!\diagram{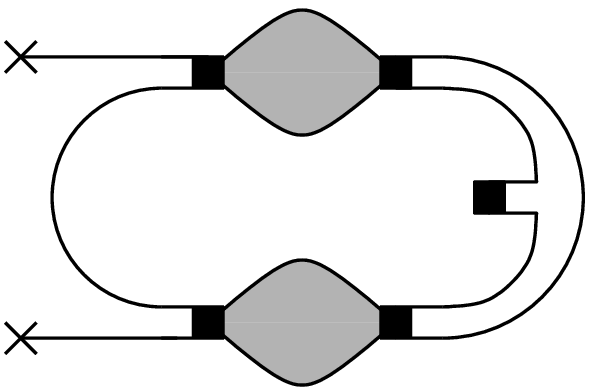}\!\!+\!\!\diagram{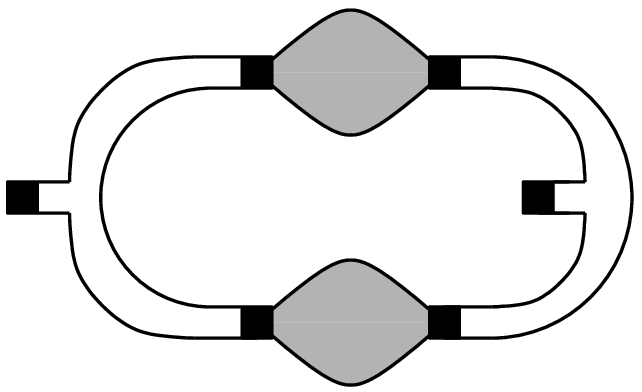}\!\!+\!\!\!\diagram{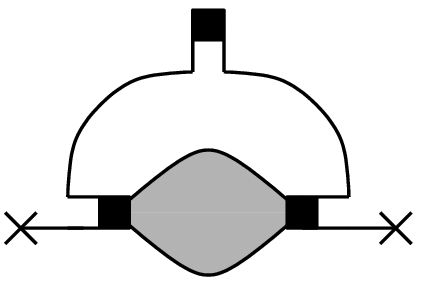}\!\!+\!\!\diagram{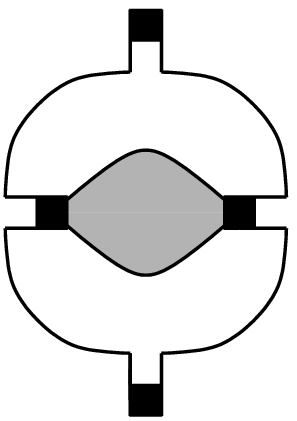}\!\! \nonumber
\\
&& +T\Big( \!\!\diagram{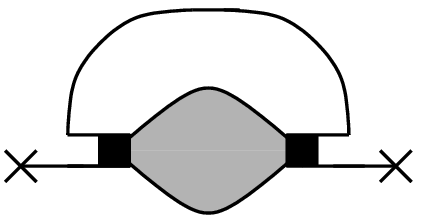} \!\!+ \!\!\diagram{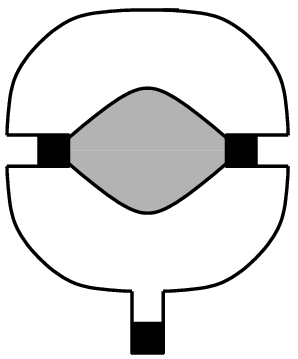} \!\!+
\!\!\diagram{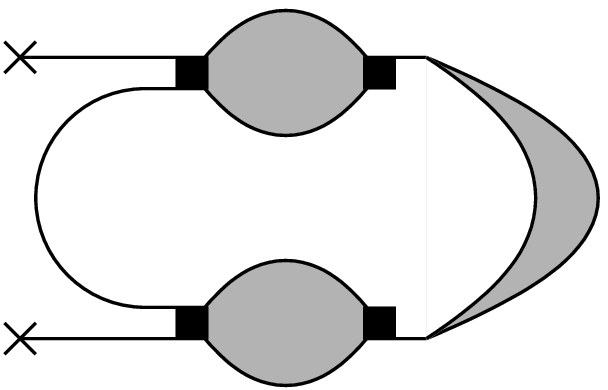} \!\!+ \!\!\diagram{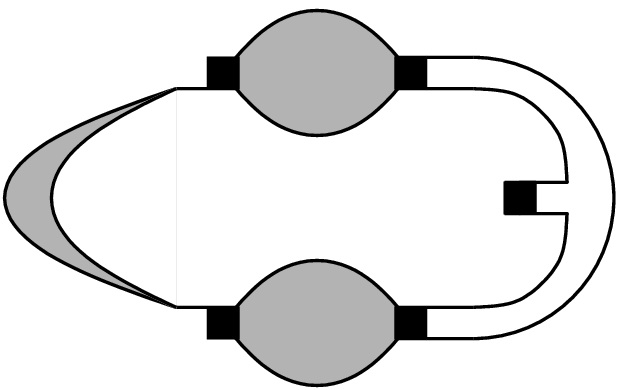}\!\!+
\!\!\diagram{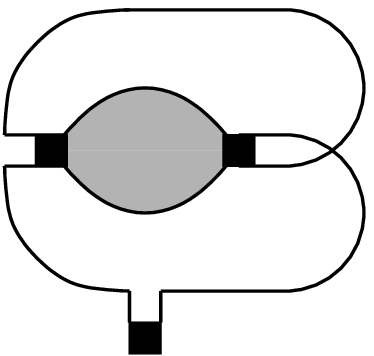} \!\!+ \!\!\diagram{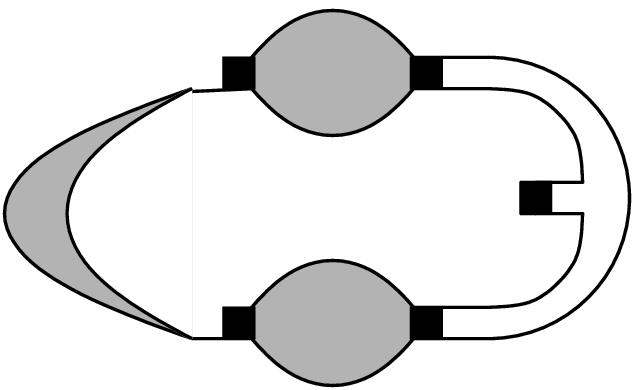}\!\! \Big)\nonumber \\
&& + T^{2}\Big( \!\!\diagram{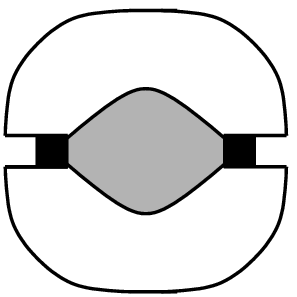} \!\!+ \!\!\diagram{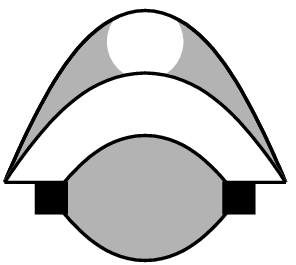}
\!\!+
\!\!\diagram{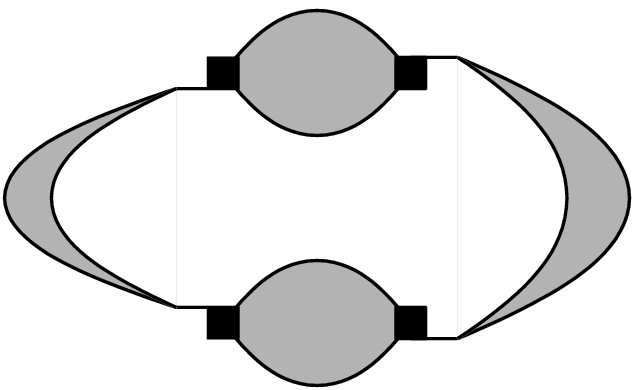} + {\cal A}^{T^{2}}\Big)\\
\diagram{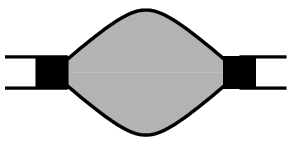}&=&B'' (\chi _{ab})\left(1-4A_{d} I_{2} (p)B'' (\chi
_{ab}) \right)^{-1}\ ,\quad  \diagram{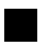}=B(\chi_{ab})\ ,
\end{eqnarray}
where the explicit expressions are given in
\cite{LeDoussalWiese2004a}.

By varying the IR-regulator, one can derive a $\beta$-function at
order $1/N$, see \cite{LeDoussalWiese2004a}. At $T=0$, it is
UV-convergent, and should allow to find a fixed point. We have been
able to do this at order $\epsilon$, showing consistency with the
1-loop result, see section \ref{s:finiteN}. Other dimensions are more
complicated.

A $\beta$-function can also be defined at finite $T$. However since
temperature is an irrelevant variable, it makes the theory
non-renormalizable, i.e.\ in order to define it, one must keep an
explicit infrared cutoff. These problems have not yet been settled.

\section{Depinning transition}\label{s:dynamics}
\begin{figure}[b]
\centerline{\fig{0.4\textwidth}{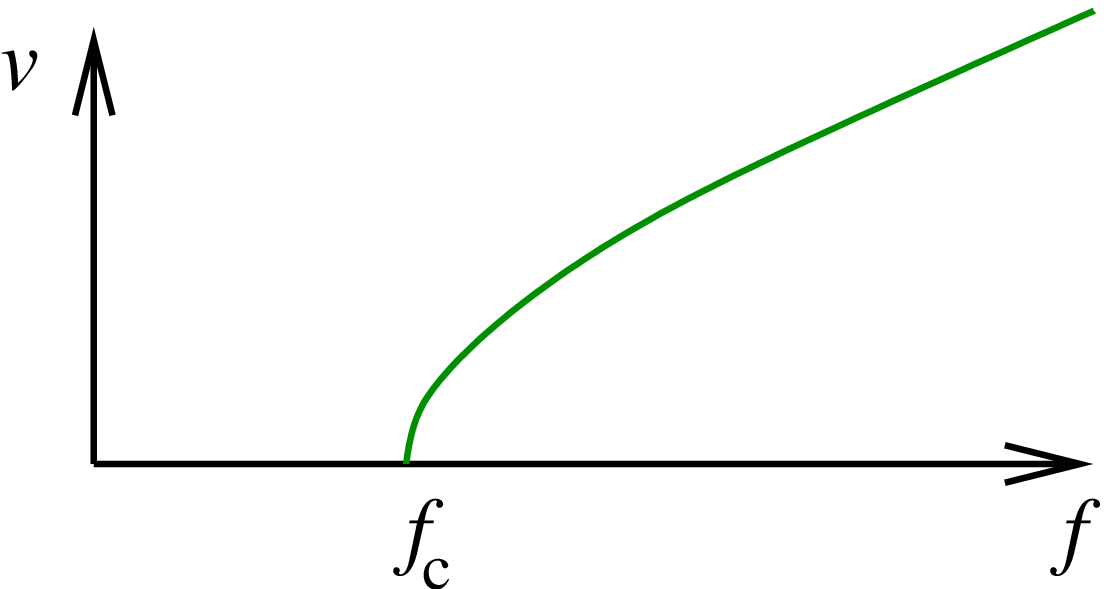}}
\Caption{Velocity of a pinned interface as a function of the applied
force. Zero force: equilibrium. $f=f_{c}$: depinning.}
\label{f:vel-force}
\end{figure}
Another important class of phenomena for elastic manifolds in disorder
is the so-called ``depinning transition'': Applying a constant force
to the elastic manifold, e.g.\ a constant magnetic field to the
ferromagnet mentioned in the introduction, the latter will only move,
if a certain critical threshold force $f_{c}$ is surpassed, see
figure \ref{f:vel-force}. (This is fortunate, since otherwise the
magnetic domain walls in the hard-disc drive onto which this article
is stored would move with the effect of deleting all information,
depriving the reader from his reading.)  At $f=f_{c}$, the so-called
depinning transition, the manifold has a distinctly different
roughness exponent $\zeta$ (see Eq.~(\ref{roughness})) from the
equilibrium ($f=0$). For $f>f_{c}$, the manifold moves, and close to
the transition, new observables and corresponding exponents appear:
\begin{itemize}
\itemsep0mm
\item  the dynamic exponent $z$ relating correlation functions in
spatial and temporal direction
$$
t\sim x^{\,z}
$$
\item a correlation length $\xi$ set by the distance to $f_{c}$
$$
\xi \sim |f-f_{c}|^{-\nu }
$$
\item  furthermore, the new exponents are not all independent, but
satisfy the following exponent relations
\cite{NattermanStepanowTangLeschhorn1992}
\begin{equation}\label{exp-relatons}
\beta =\nu (z- \zeta ) \qquad \qquad \nu =\frac{1}{2-\zeta }
\end{equation}
\end{itemize}
The equation describing the movement of the interface is
\begin{equation}\label{eq-motion}
\partial_{t} u (x,t) = (\nabla^{2}+m^{2}) u (x,t) + F (x,u (x,t)) \ ,
\qquad F (x,u)=-\partial_{u} V (x,u)
\end{equation}
This model has been treated at 1-loop order by Natterman et
al.~\cite{NattermanStepanowTangLeschhorn1992} and by Narayan and
Fisher \cite{NarayanDSFisher1993a}. The 1-loop flow-equations are
identical to those of the statics. This is surprising, since
physically, the phenomena at equilibrium and at depinning are quite
different. There is even the claim by \cite{NarayanDSFisher1993a},
that the roughness exponent in the random field universality class is
exactly $\zeta =\epsilon /3$, as it is in the equilibrium random field
class. After a long debate among numerical physicists, the issue is
today resolved: The roughness is significantly larger, and reads e.g.\
for the driven polymer $\zeta =1.25$, instead of $\zeta=1$ as
predicted in \cite{NarayanDSFisher1993a}. Clearly, a 2-loop analysis
\cite{LeDoussalWieseChauve2002} is necessary, to resolve these
issues. Such a treatment starts from the dynamic action
\begin{equation}\label{dyn-action}
{\cal S} = \int_{x,t} \tilde u (x,t) (\partial_{t}-\nabla^{2}+m^{2}) u
(x,t) +\int_{x,t,t'} \tilde u (x,t)\Delta (u (x,t)-u (x,t'))\tilde u
(x,t')\ ,
\end{equation}
where the ``response field'' $\tilde u (x,t)$ enforces the equation of
motion (\ref{eq-motion}) and
\begin{equation}\label{Delta}
\overline{F (x,u) F (x',u')} = \Delta (u-u')\delta^{d} (x-x') \equiv
-R'' (u-u') \delta^{d}(x-x')
\end{equation}
is the force-force correlator, leading to the second term in
(\ref{dyn-action}). As in the statics, one encounters terms
proportional to $\Delta' (0^{+})\equiv -R''' (0^{+})$. Here the
sign-problem can uniquely be solved by observing that the membrane
only jumps ahead,
\begin{equation}\label{jump-ahead}
t>t'\qquad \Rightarrow \qquad u (x,t)\ge u (x,t')\ .
\end{equation}
Practically this means that when evaluating diagrams containing
$\Delta (u (x,t)-u (x,t'))$, one splits them into two pieces, one with
$t<t'$ and one with $t>t'$. Both pieces are well defined, even in the
limit of $t\to t'$. As the only tread-off of this method, diagrams can
become complicated and difficult to evaluate; however they are always
well-defined.

\begin{figure}[b]
\scalebox{1.0}{
\begin{tabular}{|c|c|c|c|c|r|}
\hline
 & $d$ & $\epsilon$ & $\epsilon^2$ & estimate & simulation~~~\\
\hline
\hline
        & $3$ & 0.33 & 0.38 & 0.38$\pm$0.02 & 0.34$\pm$0.01  \\
\hline
$\zeta$ & $2$ & 0.67 & 0.86 & 0.82$\pm$0.1 & 0.75$\pm$0.02  \\
\hline
        & $1$ & 1.00 & 1.43 & 1.2$\pm$0.2 & 1.25$\pm$0.01   \\
\hline
\hline
        & $3$ & 0.89 & 0.85 & 0.84$\pm$0.01 & 0.84$\pm$0.02  \\
\hline
$\beta$ & $2$ & 0.78 & 0.62 & 0.53$\pm$0.15  & 0.64$\pm$0.02
\\
\hline
        & $1$ & 0.67 & 0.31 & 0.2$\pm$0.2 &  0.25 \dots 0.4  \\
\hline
\hline
        & $3$ & 0.58 & 0.61 &   0.62$\pm$0.01  & \\
\hline
$\nu$ & $2$ & 0.67 & 0.77 &  0.85$\pm$0.1   & 0.77$\pm$0.04  \\
\hline
        & $1$ & 0.75 & 0.98 &   1.25$\pm$0.3  & 1$\pm$0.05  \\
\hline
\end{tabular}}\hfill
%
{%
\begin{tabular}{|c|c|c|c|c|c|}
\hline
 & $\epsilon $ & $\epsilon ^{2}$ & estimate & simulation \\
\hline
\hline
$\zeta $& 0.33     &  0.47    & 0.47$\pm$0.1 & 0.39$\pm$0.002 \\
 \hline
 $\beta $  & 0.78 & 0.59 & 0.6$\pm $0.2 &0.68$\pm$0.06   \\
\hline
$z$  &0.78 &0.66 &0.7$\pm $0.1 & 0.74$\pm$0.03 \\
\hline $\nu $ & 1.33 & 1.58 & 2$\pm$0.4 &1.52$\pm $0.02 \\
\hline
\end{tabular}}
\Caption{The critical exponents at the depinning transition, for short
range elasticity (left) and for long range elasticity (right).}
\label{dyn-data}
\end{figure}
Physically, this means that we approach the depinning transition from
above. This is reflected in (\ref{jump-ahead}) by the fact that $u
(x,t)$ may remain constant; and indeed correlation-functions at the
depinning transition are completely independent of
time\cite{LeDoussalWieseChauve2002}.  On the other hand a theory for
the approach of the depinning transition from below ($f<f_{c}$) has
been elusive so far.

\begin{figure}[t]\setlength{\unitlength}{1.4mm}
\fboxsep0mm   
\psfrag{random-field-disorder-num}[][]{\small RF $m=0.071$, $L=512$}
\psfrag{random-bond-disorder-num}[][]{\small RB $m=0.071$, $L=512$}
\psfrag{y}[][]{\small $Y (z)$}
\psfrag{x}[][]{\small $z$} 
\centerline{\fig{9cm}{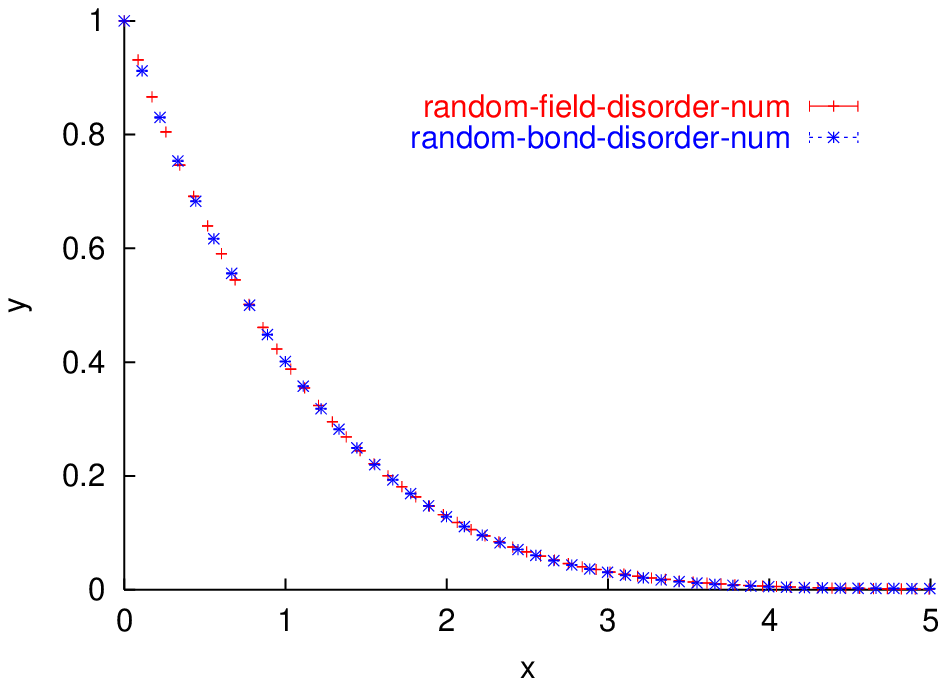}}
\Caption{Universal scaling form $Y (z)$ for $\Delta (u)$ for RB and RF
disorder. Reprinted from \cite{RossoLeDoussalWiese2006a}.} \label{f:DeltaRosso}
\end{figure}

At the depinning transition, the 2-loop functional RG reads
\cite{ChauveLeDoussalWiese2000a,LeDoussalWieseChauve2002}
\begin{eqnarray}\label{two-loop-FRG-dyn}
\partial_{\ell} R (u) \!&=&\! (\epsilon -4 \zeta)R (u)+\zeta u R' (u)
+\frac{1}{2}R'' (u)^{2} {-}R'' (u) R'' (0) \nonumber \\
&&+\frac{1}{2}\,\left[R'' (u)-R'' (0) \right] R''' (u)^{2}
~{\mbox{\bf +}}~ \frac{1}{2}\, R''' (0^{+})^{2} R''
(u)
\end{eqnarray}
First of all, note that it is a priori not clear that the functional
RG equation, which is a flow equation for $\Delta (u)=-R'' (u)$ can be
integrated to a functional RG-equation for $R (u)$. We have chosen
this representation here, in order to make the difference to the
statics evident: The only change is in the last sign on the second
line of (\ref{two-loop-FRG-dyn}). This has important consequences for
the physics: First of all, the roughness exponent $\zeta$ for the
random-field universality class changes from $\zeta
=\frac{\epsilon}{3}$ to
\begin{equation}\label{zetaRFdyn}
\zeta =\frac{\epsilon}{3} (1 +0.14331 \epsilon +\dotsb )
\end{equation}
Second, the random-bond universality class is unstable and always
renormalizes to the random-field universality class, as is physically
expected: Since the membrane only jumps ahead, it always experiences a
new disorder configuration, and there is no way to know of whether this
disorder can be derived from a potential or not. 
Generalizing the arguments of section \ref{measurecusp}, it has recently been confirmed numerically that both RB and RF disorder flow to the RF fixed point \cite{LeDoussalWiese2006a,RossoLeDoussalWiese2006a}, and that this fixed point is very close to the solution of (\ref{two-loop-FRG-dyn}), see figure \ref{f:DeltaRosso}.
This non-potentiality
is most strikingly observed in the random periodic universality class,
which is the relevant one for charge density waves. The fixed point
for a periodic disorder of period one reads (remember $\Delta (u)=-R''
(u)$)
\begin{equation}\label{rand-per-fp}
\Delta^{*} (u) =\frac{\epsilon}{36}+\frac{\epsilon^{2}}{108}
-\left(\frac{\epsilon}{6}+\frac{\epsilon^{2}}{9} \right) u (1-u)
\end{equation}
Integrating over a period, we find (suppressing in $F (x,u)$ the
dependence on the coordinate $x$ for simplicity of notation)
\begin{equation}\label{period}
\int_{0}^{1}\rmd u \, \Delta^{*} (u) \equiv \int_{0}^{1}\rmd u\
\overline{F (u) F (u')}= -\frac{\epsilon^{2}}{108}\ .
\end{equation}
In an equilibrium situation, this correlator would vanish, since
potentiality requires $\int_0^{1}\rmd u\, F (u)\equiv 0$. Here, there
are non-trivial contributions at 2-loop order (order $\epsilon^{2}$),
violating this condition and rendering the problem non-potential. This
same mechanism is also responsible for the violation of the conjecture
$\zeta =\frac{\epsilon}{3}$, which could be proven on the assumption
that the problem remains potential under renormalization. Let us
stress that the breaking of potentiality under renormalization is a
quite novel observation here.

The other critical exponents mentioned above can also be
calculated. The dynamical exponent $z$ (for RF-disorder) reads
\cite{ChauveLeDoussalWiese2000a,LeDoussalWieseChauve2002}
\begin{equation}\label{zdyn}
z=2-\frac{2}{9}\epsilon -0.04321\epsilon^{2} + \dotsb
\end{equation}
All other exponents are related via the relation (\ref{exp-relatons}).
That the method works well even quantitatively can be inferred from
figure \ref{dyn-data}.

\section{Supersymmetry}\label{a5} The use of $n$ replicas in the limit
$n\to 0$ to describe disordered systems is often criticized for a lack
of rigor. It is argued that instead one should use a supersymmetric
formulation. Such a formulation is indeed possible, both for the
statics as discussed in \cite{Wiese2004}, as for the dynamics, which
we will discuss below. Following \cite{ParisiSourlas1979}, one groups
the field $u (x)$, a bosonic auxiliary field $\tilde u (x)$ and two
Grassmanian fields $\psi (x)$ and $\bar \psi (x)$ into a superfield $U
(x,\bar \Theta , \Theta )$:
\begin{equation}\label{superfielddef}
U (x,\bar \Theta ,\Theta) = u (x)+ \bar \Theta \psi (x)+\bar \psi (x)
\Theta + \Theta \bar \Theta \tilde u (x)
\ .
\end{equation}
The action of the supersymmetric theory is
\begin{equation}\label{17.2}
{\cal S}_{\mathrm{Susy}}= \int \rmd \Theta \rmd \bar \Theta\int_{x} U
(x,\bar \Theta ,\Theta) (\Delta_{s}) U (x,\bar \Theta ,\Theta)\ ,
\qquad \Delta_{s} := \nabla^{2}-\Delta (0) \frac{\partial}{\partial
\bar \Theta}\frac{\partial}{\partial \Theta}
\end{equation}
It is invariant under the action of the supergenerators
\begin{equation}\label{17.3}
Q := x \frac{\partial}{\partial \Theta}-\frac{2}{\Delta (0)} \bar
\Theta \nabla \ , \qquad \bar Q:=x \frac{\partial}{\partial \bar
\Theta}+\frac{2}{\Delta (0)} \Theta \nabla\ .
\end{equation}
What do the fields mean?  Upon integrating over $\bar \Theta$ and
$\Theta$ before averaging over disorder, one would obtain two terms,
$\sim \int_{x} \tilde u (x) \frac{\delta {\cal H}}{\delta u (x)}$,
i.e.\ the bosonic auxiliary field $\tilde u (x)$ enforces
$\frac{\delta {H}}{\delta u (x)}=0$, and a second term, bilenear in
$\bar \psi$ and $\psi$, $\sim \int_{x}\bar \psi \frac{\delta^{2}{\cal
H}}{\delta u^{2}}\psi$ which ensures that the partition function is
one.  (\ref{17.2}) is nothing but the dimensional reduction result
(\ref{zetaDR}) in super-symmetric disguise. What went wrong?  Missing
is the renormalization of $R (u)$ itself, which in the FRG approach
leads to a flow of $\Delta (0)\equiv -R'' (0)$. In order to capture this,
one has to look at the supersymmetric action of at least two copies:
\begin{equation}\label{17.4}
{\cal S}[U_{a}]= \sum_{a}\int_{\Theta, \bar \Theta }\int_{x}
U_{a} (x,\bar \Theta ,\Theta) (\Delta_{s}) U_{a} (x,\bar \Theta
,\Theta) -\half  \sum_{a\neq b} \int_{x}\int_{\bar \Theta ,\Theta} \int_{
 \bar \Theta', \Theta'} R (U_{a} (x,\bar \Theta ,\Theta)-U_{b}
(x,\bar \Theta ',\Theta' ))
\ .
\end{equation}
Formally, we have again introduced $n$ replicas, but we do not take
the limit of $n\to 0$; so criticism of the latter limit can not be
applied here. After some slightly cumbersome calculations one
reproduces the functional RG $\beta$-function at 1-loop
(\ref{1loopRG}). (Higher orders are also possible and the SUSY method
is actually helpful \cite{Wiese2004}.) At the Larkin-length, where the
functional RG produces a cusp, the flow of $\Delta (0)$ becomes
non-trivial, given in (\ref{R2of0after}). Then the parameter $\Delta
(0)$ in the supersymmetry generators (\ref{17.3}) is no longer a
constant, and supersymmetry breaks down. This is, as was discussed
in section (\ref{s:RSB}), also the onset of replica-symmetry breaking
in the gaussian variational ansatz, valid at large $N$.

Another way to introduce a supersymmetric formulation proceeds via the
super-symmetric representation of a stochastic equation of motion
\cite{Zinn}; a method e.g.\ used in \cite{Kurchan1992} to study
spin-glasses. The action then changes to
\begin{equation}\label{17.5}
{\cal S}[U]= \int_{xt}\int_{\Theta, \bar \Theta }
U (x,\bar \Theta ,\Theta,t) (\Delta_{d}) U(x,\bar \Theta
,\Theta,t) -\half  \int_{xtt'}\int_{\bar \Theta ,\Theta} \int_{
 \bar \Theta', \Theta'} R (U (x,\bar \Theta ,\Theta,t)-U
(x,\bar \Theta ',\Theta' ,t'))
\ .
\end{equation}
\begin{equation}\label{a9}
\Delta_{d} = \nabla^{2} +  \bar D D \ , \qquad \bar D =
\frac{\partial}{\partial \Theta}\ ,\qquad  D= \frac{\partial}{\partial
\bar \Theta} -\Theta \frac{\partial}{\partial t}
\end{equation}
and is invariant under the action of the super-generators $Q:=
\frac{\partial}{\partial \bar \Theta} $ and $\bar Q :=
\frac{\partial}{\partial \Theta} +\bar \Theta \frac{\partial}{\partial
t}$, since $\left\{Q,D \right\}=\left\{Q,\bar D \right\}=\left\{\bar
Q,D \right\} =\left\{\bar Q,\bar D \right\}=0$. Different replicas now
become different times, but second cumulant still means bilocal in
$\Theta$. However the procedure is not much different from a pure
Langevin equation of motion, as in (\ref{eq-motion}); in the latter
equation It\^o discretization is already sufficient to ensure that the
partition function is 1. The main advantage is the possibility to
change the discretization procedure from It\^o over mid-point to
Stratonovich without having to add additional terms. In this case,
supersymmetry breaking means that the system falls out of equlibrium,
i.e.\ the fluctuation-dissipation theorem (which is a consequence of
one of the supersymmetry generators \cite{Zinn}) breaks down
\cite{Kurchan1992}.

\section{Random Field Magnets}\label{a6}
Another domain of application of the Functional RG is spin models in
a random field. The model usually studied is:
\begin{eqnarray}
{\cal H} = \int \rmd^d x\, \half (\nabla \vec S)^2 + \vec h(x) \cdot \vec
S(x) \ , \label{rf}
\end{eqnarray}
where $\vec S(x)$ is a unit vector  with
$N$-components, and $\vec S(x)^2=1$. This is the so-called $O(N)$ sigma model, to which has
been added a random field, which can be taken gaussian
$\overline{h_i(x) h_j(x')} = \sigma \delta_{ij} \delta^d(x-x')$. In
the absence of disorder the model has a ferromagnetic phase for
$T<T_{\mathrm{f}}$ and a paramagnetic phase above $T_{\mathrm{f}}$.
The lower critical dimension is $d=2$ for any $N \geq 2$, meaning that
below $d=2$ no ordered phase exists.  In $d=2$ solely a paramagnetic
phase exists for $N>2$; for $N=2$, the XY model, quasi long-range
order exists at low temperature, with $\overline{\vec S(x) \vec
S(x')}$ decaying as a power law of $x-x'$.

Here we study the model directly at $T=0$. The dimensional reduction
theorem in section \ref{dimred}, which also holds for random field
magnets, would indicate that the effect of a quenched random field in
dimension $d$ is similar to the one of a temperature $T \sim \sigma$
for a pure model in dimension $d-2$. Hence one would expect a
transition from a ferromagnetic to a disordered phase at $\sigma_c$ as
the disorder increases in any dimension $d>4$, and no order at all for
$d<4$ and $N \geq 2$. This however is again incorrect, as can be seen
using FRG.

It was noticed by Fisher \cite{Fisher1985b} that an infinity of
relevant operators are generated. These operators, which correspond to
an infinite set of random anisotropies, are irrelevant by naive power
counting near $d=6$ \cite{Feldman2000,Feldman2000b}. $d=6$ is the
naive upper critical dimension (corresponding to $d=4$ for the pure
$O(N)$ model) as indicated by dimensional reduction; so many earlier
studies concentrated on $d$ around $6$. Because of these operators the
theory is however hard to control there. It has been shown
\cite{Fisher1985b,Feldman2000,Feldman2000b} that it can be controlled
using 1-loop FRG near $d=4$ instead, which is the naive lower critical
dimension. Recently this was extended to two loops
\cite{LeDoussalWiese2005b}.

The 1-loop FRG studies directly the model with all the operators
which are marginal in $d=4$, of action most easily expressed
directly in replicated form:
\begin{eqnarray} \label{action}
 {\cal S} = \int \rmd^d x &\Big[&\! \frac{1}{2 T} \sum_a [ (\nabla
\vec S_a)^2 ] - \frac{1}{2 T^2} \sum_{a b} \hat R(\vec S_a \vec S_b)
\Big]\ ,
\end{eqnarray}
The function $\hat R(z)$ parameterizes the disorder. Since the vectors
are of unit norm, $z=\cos \phi$ lies in the interval $[-1,1]$.  One
can also use the parametrization in terms of the variable $\phi$ which
is the angle between the two replicas, and define $R(\phi)=\hat
R(z=\cos \phi)$. The original model (\ref{rf}) corresponds to $\hat
R(z) \sim \sigma z$. It does not remain of this form under RG, in fact
again a cusp will develop near $z=1$. The FRG flow equation has been
calculated up to two loops, i.e.\ $R^2$ (one loop)
\cite{Fisher1985b,Feldman2000,Feldman2000b} and $R^3$ (two loops)
\cite{LeDoussalWiese2005b}\footnote{These results were confirmed in
\cite{TarjusTissier2005} (for the normal terms not proportional to $R'''
(0^{+})$) and \cite{TarjusTissier2006} (with one proposition for the
anomalous terms).}:
\begin{eqnarray}
\partial_{\ell} R (\phi ) &=& \epsilon R (\phi )+ \half R''
(\phi)^2-R''(0)R''(\phi) + (N{-}2)\left[\frac 1 2
\frac{R'(\phi)^2}{\sin^2 \phi }-
 \cot \phi R'(\phi)R''(0)\right] \nonumber \\
&& +  \half (R''(\phi)-R''(0) ) R'''(\phi)^2 +
 (N{-}2) \bigg[ \frac{\cot \phi}{\sin^4 \phi} R'(\phi)^3
- \frac{5+ \cos 2 \phi}{4 \sin^4 \phi} R'(\phi)^2 R''(\phi)
\nonumber \\&& + \frac{1}{2 \sin^2 \phi} R''(\phi)^3  - \frac{1}{4
\sin^4 \phi} R''(0) \Big( 2 (2 + \cos 2 \phi) R'(\phi)^2 - 6 \sin 2
\phi
R'(\phi) R''(\phi)  \nonumber \\
&&
+(5+ \cos 2 \phi) \sin^2 \phi R''(\phi)^2 \Big) \bigg] \nonumber \\
&& - \frac{N{+}2} 8  R'''(0^+)^2 R'' (\phi ) - \frac{N{-}2}{4} \cot
\phi R'''(0^+)^2 R' (\phi )
\nonumber \\
&&  - 2 (N{-}2) \Big[R'' (0) - R'' (0)^{2} + \gamma_{a} R'''
(0^+)^{2} \Big]
   R (\phi ) \qquad  \label{beta}
\end{eqnarray}
The constant $\gamma_a$ is discussed in \cite{LeDoussalWiese2005b};
the last factor proportional to $R (\phi )$ takes into account the
renormalization of temperature, a specific feature absent in the
manifold problem.  The full analysis of this equation is quite
involved. The 1-loop part already shows interesting
features. For $N=2$, the fixed point was studied in
\cite{GiamarchiLeDoussal1995}, and corresponds to the Bragg-glass
phase of the XY model with quasi-long range order obtained in a
$d=4-\epsilon$ expansion below $d=4$. Hence for $N=2$ the lower
critical dimension is $d_{\mathrm{lc}} < 4$ and conjectured to be
$d_{\mathrm{lc}} < 3$ in \cite{GiamarchiLeDoussal1995}. On the other
hand Feldman \cite{Feldman2000,Feldman2000b} found that for $N=3,
4,\dots$ there is a fixed point for $d=4+\epsilon$ for $d>4$.  This
fixed point has exactly one unstable direction, hence was conjectured
to correspond to the ferromagnetic-to-disorder transition. The
situation at one loop is thus rather strange: For $N=2$, only a stable
FP which describes a {\em unique} phase exists, while for $N=3$ only
an unstable FP exists, describing the transition between two
phases. The question is: Where does the disordered phase go as $N$
decreases?  These results cannot be reconciled within one loop and
require the full 2-loop analysis of the above FRG equation.

\begin{figure}[t]
\Fig{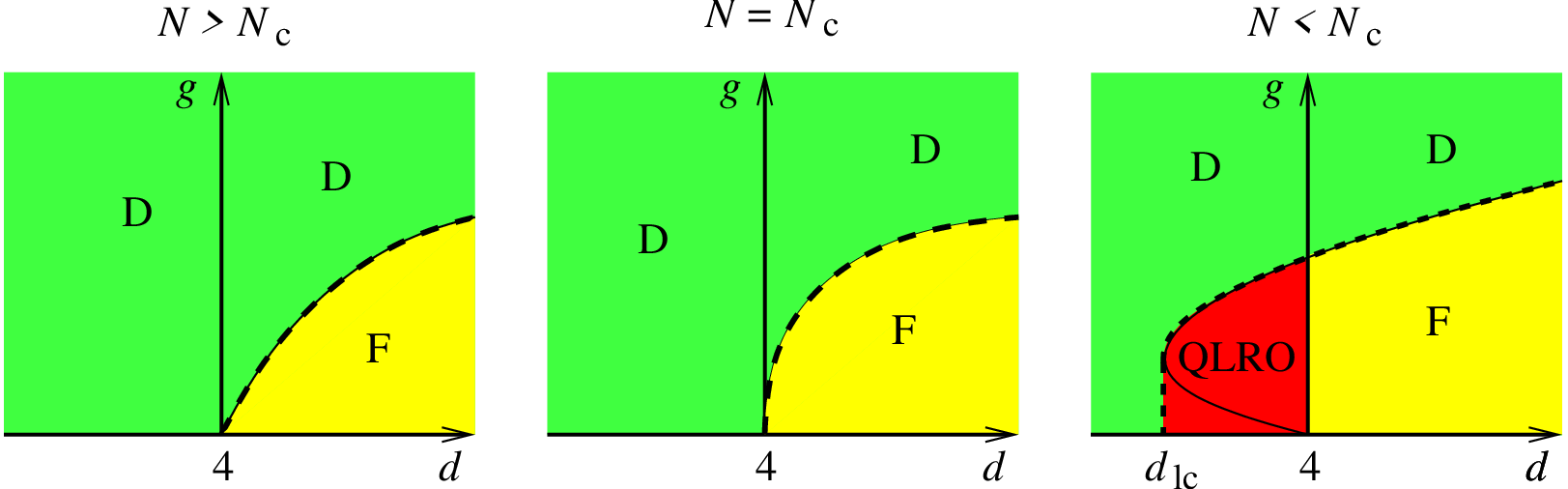}
\Caption{Phase diagram of the RF non-linear sigma model. D $=$ disordered, F $=$ ferromagnetic,
QLRO $=$ quasi long-range order. Reprinted from \cite{LeDoussalWiese2005b}.}
\label{f:phases}
\end{figure}
The complete analysis \cite{LeDoussalWiese2005b} shows that there is a
critical value of $N$, $N_c=2.8347408$, below which the lower critical
dimension $d_{\mathrm{lc}}$ of the quasi-ordered phase plunges below
$d=4$. Hence there are now two fixed points below $d=4$. For $N>N_c$ a
ferromagnetic phase exists with lower critical dimension
$d_{\mathrm{lc}}=4$. For $N<N_c$ one finds an expansion:
\begin{equation}\label{expansiondc}
d_{\mathrm{lc}}^{\mathrm{RF}} = 4-\epsilon_{c}\approx 4 - 0.1268
(N-N_{c})^{2}+ O ( (N-N_{c})^{3})\ . 
\end{equation}
One can also compute the exponents of the correlation function
\begin{eqnarray}\label{a10}
 \overline{S_q S_{-q} } \sim q^{-4 + \bar \eta}\ ,
\end{eqnarray}
and once the fixed point is known, $\bar \eta$ is given by $\bar
\eta=\epsilon- (N-1) R''(0) + {\textstyle \frac{3 N-2}{8}}
R'''(0^+)^2$. There is a similar exponent for the connected thermal
correlation $\eta$. Another fixed point describing magnets with 
random anisotropies (i.e.\ disorder coupling linearly to $S_i(x) S_j(x)$)
is studied in \cite{Feldman2000,Feldman2000b,LeDoussalWiese2005b, KuehnelLeDoussalWieseUnbublished}.

In this context, the existence of a quasi-ordered phase for the
random-field XY model in $d=3$ (the scalar version of the Bragg glas) 
has been questioned \cite{TarjusTissier2005}. Corrections in
(\ref{expansiondc}) seem to be small and at first sight exclude the
quasi-ordered phase in $d=3$. This should however be taken with a
(large) grain of salt \cite{LeDoussalIHP2006}. 
Indeed the above model does not even contain topological defects
(i.e.\ vortices) as it was directly derived in the continuum. In the
absence of topological defects it is believed that the lower critical
dimension is $d_{\mathrm{lc}}=2$ (with logarithmic corrections
there). Hence the above series should converge to that value for
$N=2$, indicating higher order corrections to
(\ref{expansiondc}). Another analysis \cite{TarjusTissier2004} based
on a FRG on the soft-spin model, which may be able to capture
vortices, indicates $d_{\mathrm{lc}}(N=2)> 3$. Unfortunately, it uses
a truncation of FRG which cannot be controlled perturbatively, and as
a result, does not match the 2-loop result. It would be interesting to
construct a better approximation which predicts accurately at which
dimension the soft and hard spin model differ in their lower critical
dimensions, probably when vortices become unbound due to
disorder. 


\section{More universal distributions}\label{s:distribution}
\begin{figure}[t] \centerline{\fig{0.5\textwidth}{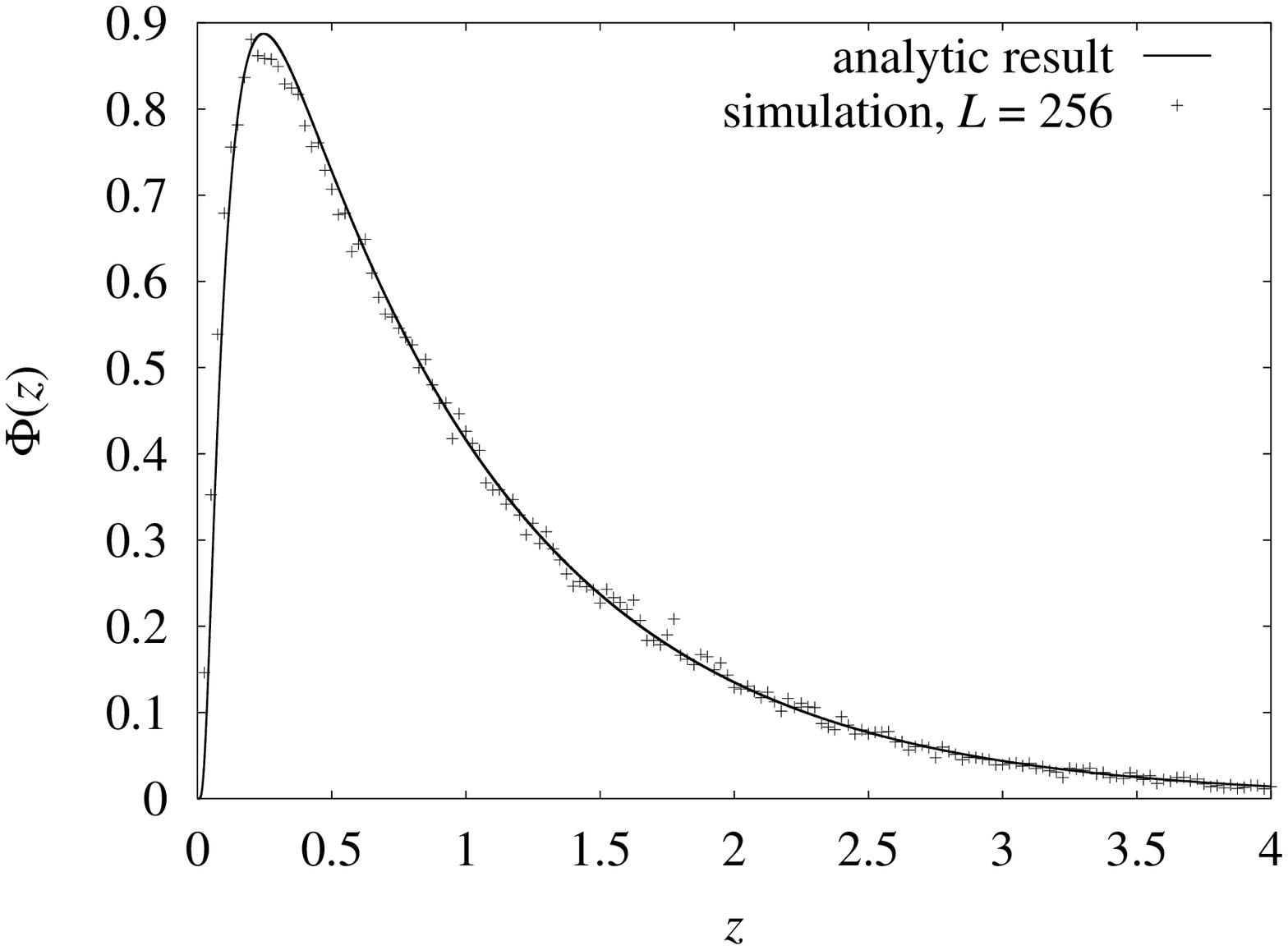}}
\Caption{Scaling function $\Phi(z)$ for the ($1+1$)--dimensional
harmonic model, compared to the Gaussian approximation for
$\zeta=1.25$. Data from
\protect\cite{RossoKrauthLeDoussalVannimenusWiese2003}.}
\label{f:Dhm}
\end{figure}
As we have already seen, 
exponents are not the only interesting quantities: In experiments and
simulations, often whole distributions can be measured, as e.g.\ the
universal width distribution of an interface that we have computed at
depinningc
\cite{RossoKrauthLeDoussalVannimenusWiese2003,LeDoussalWiese2003a}. Be
$\left< u \right>$ the average position of an interface for a {\em
given} disorder configuration, then the spatially averaged width
\begin{eqnarray}\label{w2}
w^{2}:= \frac{1}{L^{d}}\int_{x}\left(u (x)-\left< u \right> \right)^{2}
\end{eqnarray}
is a random variable, and we can try to calculate and measure its
distribution $P (w^{2})$. The rescaled function $\Phi (z)$, defined by
\begin{equation}\label{Phi}
P (w^{2})={1}/{\overline{w^{2}}}\,\Phi
\left({w^{2}}/{\overline{w^{2}}} \right)
\end{equation}
will be universal, i.e.~independent of microscopic details and the
size of the system.

Supposing all correlations to be Gaussian, $\Phi (z)$ can be
calculated analytically. It depends on two parameters, the roughness
exponent $\zeta$ and the dimension $d$. Numerical simulations
displayed on figure \ref{f:Dhm} show spectacular agreement between
analytical and numerical results. As expected, the Gaussian
approximation is not exact, but to see deviations in a simulation,
about $10^{5}$ samples have to be used. Analytically, corrections can
be calculated: They are of order $R''' (0^{+})^{4}$ and
small. Physically, the distribution is narrower than a Gaussian.

There are more observables of which distributions have been calculated
within FRG, or measured in simulations. Let us mention fluctuations of
the elastic energy \cite{FedorenkoStepanow2003}, and of the depinning
force \cite{FedorenkoLeDoussalWiese2006,BolechRosso2004}.


\section{Anisotropic depinning, directed percolation, branching and
all that}\label{s:anisotopic}
\begin{figure}[b] \centerline{\fig{2cm}{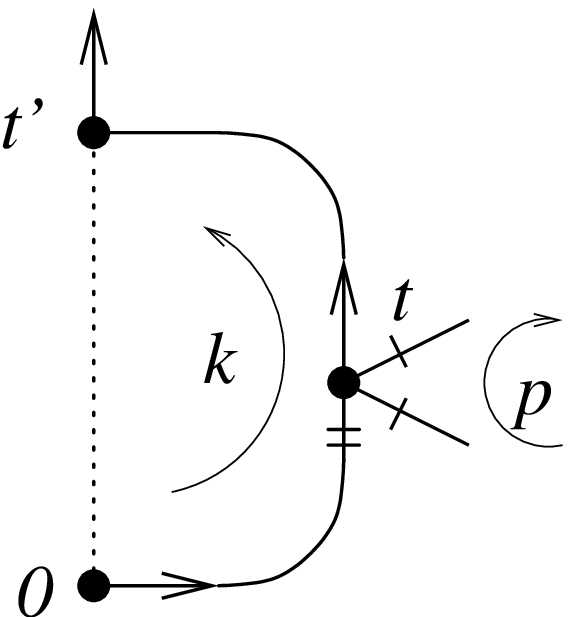}} \Caption{{The
diagram generating the irreversible nonlinear KPZ term with one
disorder vertex and one $c_4$ vertex (the bars denote spatial
derivatives).}}  \label{fig1.a}
\end{figure}
We have discussed in section \ref{s:dynamics} isotopic depinning,
which as the name suggests is a situation, where the system is
invariant under a tilt. This isotropy can be broken through an
additional anharmonic elasticity
\begin{equation}\label{Eanharm}
E_{\mathrm{elastic}}= \int_{x} \frac{1}{2}\left[\nabla
u (x)\right]^{2}+c_{4} \left[\nabla u (x)\right]^{4}\ ,
\end{equation}
leading to a drastically different universality class, the so-called
anisotropic depinning universality class, as found recently in
numerical simulations \cite{RossoKrauth2001b}. It has been observed in
simulations \cite{AmaralBarabasiStanley1994,TangKardarDhar1995}, that
the drift-velocity of an interface is increased, which can be
interpreted as a tilt-dependent term, leading to the equation of motion
in the form
\begin{equation}\label{lf28}
 \partial_t u (x,t)= \nabla^2 u (x,t) + \lambda \left[ \nabla u
(x,t)\right]^2+ F(x,u (x,t) ) + f\ .
\end{equation}
However it was for a long time unclear, how this new term
(proportional to $\lambda$), which usually is referred to as a
KPZ-term, is generated, especially in the limit of {\em vanishing}
drift-velocity. In  \cite{LeDoussalWiese2002a} we have shown that this
is possible in a non-analytic theory, due to the diagram given in
figure \ref{fig1.a}.

For anisotropic depinning, numerical simulations based on cellular
automaton models which are believed to be in the same universality
class
\cite{TangLeschhorn1992,BuldyrevBarabasiCasertaHavlinStanleyVicsek1992},
indicate a roughness exponent $\zeta \approx 0.63$ in $d=1$ and $\zeta
\approx 0.48$ in $d=2$.  On a phenomenological level it has been
argued
\cite{TangLeschhorn1992,BuldyrevBarabasiCasertaHavlinStanleyVicsek1992,GlotzerGyureSciortinoConiglioStanley1994}
that configurations at depinning can be mapped onto directed
percolation in $d=1+1$ dimensions, which yields indeed a roughness
exponent $\zeta_{\mathrm{DP}}= \nu_\perp/\nu_{\|} = 0.630 \pm 0.001$,
and it would be intriguing to understand this from a systematic field
theory.

This theory was developed in \cite{LeDoussalWiese2002a}, and we review
the main results here. A strong simplification is obtained by going
to the Cole-Hopf transformed fields
\begin{equation}\label{lf29}
Z ( {x,t}) := \rme^{ \lambda u(x,t)} \qquad \Leftrightarrow \qquad
u(x,t) = \frac{\ln ( Z(x,t))}{ \lambda } \ .
\end{equation}
The equation of motion becomes after multiplying with $ \lambda
Z(x,t)$ (dropping the term proportional to $f$)
\begin{equation}\label{lf30}
   \partial_t Z(x,t) = \nabla^2 Z(x,t) +{{\lambda} }
F\left(x,\frac{\ln (Z(x,t))}{{\lambda} } \right) Z(x,t)
\end{equation}
and the dynamical action (after averaging over disorder)
\begin{equation}\label{cole}
{\cal S} = \int_{xt}\tilde {Z}(x,t)\left( \partial_{t}-\nabla^{2}
\right) Z(x,t) -\frac{ \lambda^{2} }{2 } \int_{xtt'} \tilde {Z}(x,t)
{Z}(x,t) \, \Delta\! \left( \frac{\ln Z(x,t)-\ln Z(x,t')}{ \lambda
}\right)\tilde {Z}(x,t')Z(x,t')
\end{equation}
This leads  to the FRG flow equation at 1-loop order
\begin{eqnarray}
\partial _{\ell} \Delta (u) &=& (\epsilon -2\zeta )
\Delta (u) + \zeta u
 \Delta' (u)  -\Delta'' (u)\left(
\Delta (u)-\Delta (0) \right) - \Delta' (u)^2\nn \\
&&+2 \lambda \Delta (u) \Delta' (0^{+}) +2 \lambda ^{2}\left(\Delta
(u)^{2} +\Delta (u)\Delta (0)\right) \label{beta-2}
\end{eqnarray}
The first line is indeed equivalent to (\ref{1loopRG}) using $\Delta
(u)=-R'' (u)$. The second line is new and contains the terms induced
by the KPZ term, i.e.\ the term proportional to $\lambda$ in
(\ref{lf28}).

Equation (\ref{beta-2}) possesses the following remarkable property:
{\em A three parameter subspace of exponential functions forms an
exactly invariant subspace.}  Even more strikingly, this is true {\it
to all orders} in perturbation theory \cite{LeDoussalWiese2002a}! The
subspace in question is ($0\le u\le 1/\lambda $)
\begin{equation}\label{lf80}
\Delta(u) = \frac{\epsilon }{ \lambda^2} \left(a + b\, \rme^{- \lambda
u} + c\, \rme^{\lambda u}\right)
\end{equation}
\begin{figure*}[!t] \centerline{\fig{.5\textwidth}{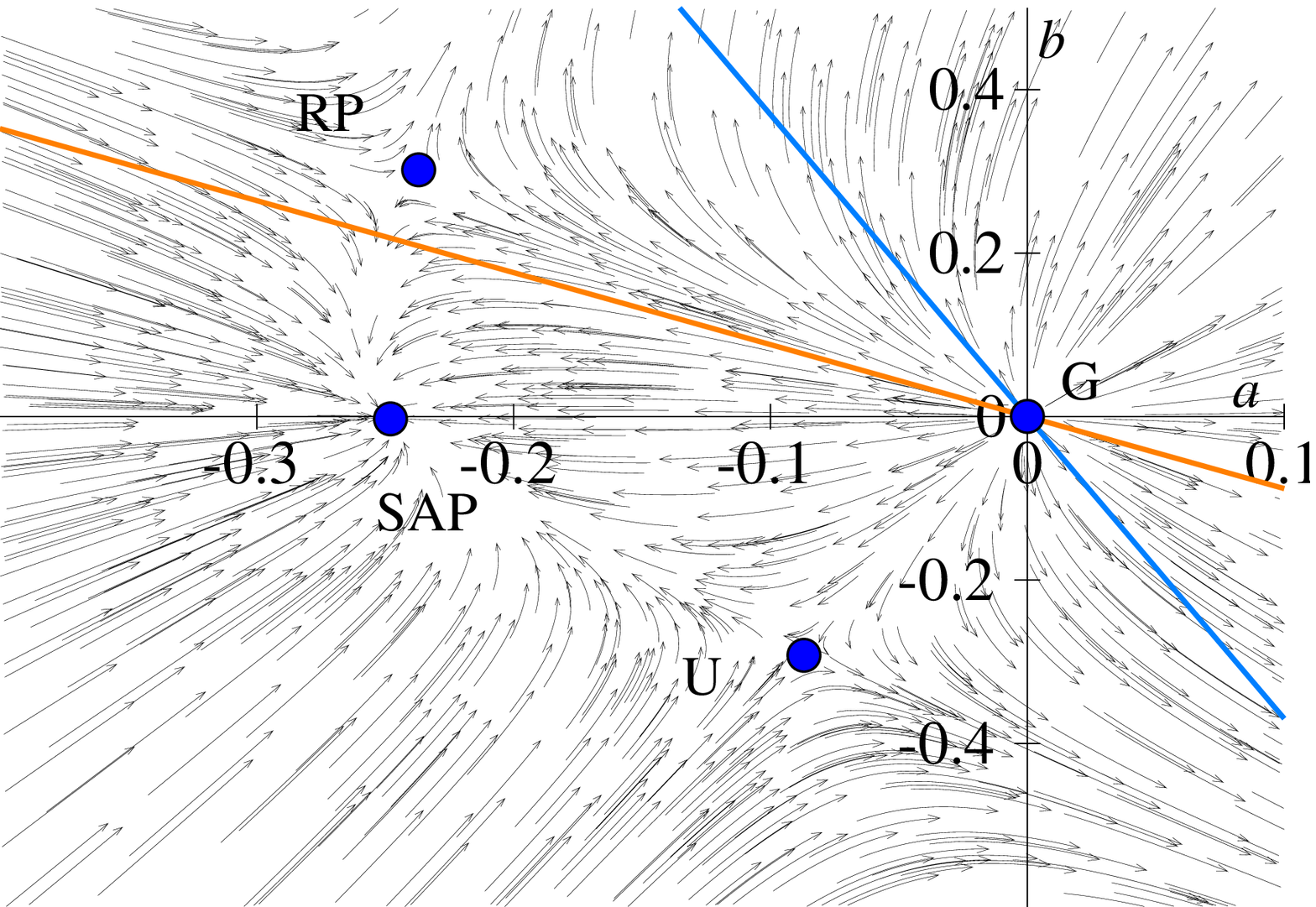}}
\Caption{Fixed point structure for $\lambda=2$, which is a typical
value. The ratio $c/b$ is not renormalized, see
(\ref{lf43})-(\ref{lf44}), such that $c/b$ is a parameter, fixed by
the boundary conditions, especially $\lambda $. The fixed points are
Gaussian {\tt G}, Random Periodic {\tt RP} (the generalization of the
RP fixed point for $\lambda =0$), Self-Avoiding Polymers {\tt SAP},
and Unphysical {\tt U}.}  \label{lambda-flow}
\end{figure*}%
The FRG-flow (\ref{beta-2}) closes in this subspace,
leading to the simpler 3-dimensional flow:
\begin{eqnarray}
 \partial_{\ell} a &=& a + 4 a^2 + 4 a c + 4 b c \label{lf42}\\
 \partial_{\ell} b &=& b (1 + 6 a  + b + 5 c ) \label{lf43}\\
 \partial_{\ell} c &=& c (1 + 6 a  + b + 5 c )\label{lf44}
\end{eqnarray} This flow has a couple of fixed points, given on figure
\ref{lambda-flow}. They describe different physical situations. The
only globally attractive fixed point is {\tt SAP}, describing
self-avoiding polymers. This fixed point is not attainable from the
physically relevant initial conditions, which lie (as fixed point {\tt
RP}) between the two separatrices given on figure \ref{lambda-flow}.
All other fixed points are cross-over fixed points.

\begin{figure}[b] \centerline{\fig{0.5\textwidth}{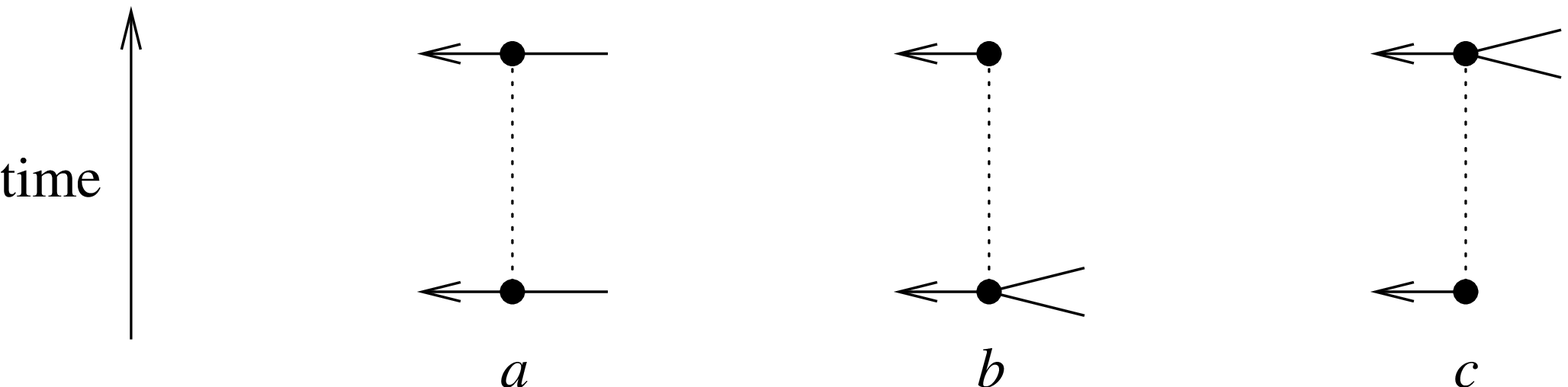}}
\Caption{The three vertices proportional to $a$, $b$ and $c$ in
equation (\ref{c2}).}  \label{f:branch}
\end{figure}In the Cole-Hopf representation, it is easy to see why the
exponential manifold is preserved to all orders. Let us insert
(\ref{lf80}) in (\ref{cole}).  The complicated functional disorder
takes a very simple polynomial form \cite{LeDoussalWiese2002a}.
\begin{equation}\label{c2}
{\cal S}=\int_{xt}\tilde {Z}(x,t)\left(
\partial_{t}-\nabla^{2}  \right) Z(x,t)- \int_{x}\int_{t<t'}
\tilde {Z}(x,t)\tilde {Z}(x,t') \left(a
Z(x,t)Z(x,t')+bZ(x,t)^{2}+cZ(x,t')^{2} \right)\ .
\end{equation}
The vertices are plotted on figure \ref{f:branch}. It is intriguing to
interpret them as particle interaction ($a$) and as different
branching processes ($b$ and $c$): $Z$ destroys a particle and $\tilde
Z$ creates one. Vertex $b$ can e.g.\ be interpreted as two particles
coming together, annihilating one, except that the annihilated
particle is created again in the future. However, if the annihilation
process is strong enough, the reappearance of particles may not play a
role, such that the interpretation as particle annihilation or
equivalently directed percolation is indeed justified.

One caveat is in order, namely that the fixed points described above,
are all transition fixed point, and nothing can a priori be said about
the strong coupling regime. However this is the regime seen in
numerical simulations, for which the conjecture about the equivalence
to directed percolation has been proposed. Thus albeit intriguing, the
above theory is only the starting point for a more complete
understanding of anisotropic depinning. Probably, one needs another
level of FRG, so as standard FRG is able to treat directed polymers,
or equivalently the KPZ-equation in the absence of disorder.

\section{Problems not treated in these notes\dots and perspectives}\label{perspectives}
Problems not treated in these notes are too numerous to list. Let us just mention some: 
\cite{FedorenkoStepanow2002} consider depinning at the upper critical dimension. In \cite{LeDoussalWiese2006a}, the crossover from short-ranged to long-ranged correlated disorder is treated. 
Our techniques can  be applied to the statics at 3-loop order
\cite{LeDoussalWiesePREPb}.  But many
questions remain open. Some have already been raised in these
notes, another is whether FRG can also be applied to other systems, as e.g.\ spinglasses or window glasses. Can FRG be used as a  tool to go beyond
mean-field or mode-coupling theories? Another open issue is the applicability 
of FRG beyond the elastic limit, i.e.\ to systems with overhangs and topological defects,  non-liear elasticity \cite{LeDoussalWieseRaphaelGolestanian2004}, or to more general fractal curves  than (directed) interfaces. 
For random periodic disorder in  $d=2$, temperature is marginal, and a freezing transition can be discussed (see e.g.\ \cite{CarpentierLeDoussal1998, SchehrLeDoussal2006}). It would be interesting to connect this to methods of conformal field theory and stochastic L\"owner evolution. 
We
have to leave these problems for future research and as a challenge for
the reader to plunge deeper into the mysteries of functional
renormalization.

\appendix \section{Derivation of the functional RG
equations}\label{app:deriveRG} In section \ref{Larkin}, we had seen
that 4 is the upper critical dimension. As for standard critical
phenomena \cite{Zinn}, we now construct an $\epsilon= (
4-d)$-expansion.  Taking the dimensional reduction result
(\ref{zetaDR}) in $d=4$ dimensions tells us that the field $u$ is
dimensionless. Thus, the width $\sigma = -R''(0)$ of the disorder is
not the only relevant coupling at small $\epsilon$, but any function
of $u$ has the same scaling dimension in the limit of $\epsilon=0$,
and might thus equivalently contribute. The natural conclusion is to
follow the full function $R(u)$ under renormalization, instead of just
its second moment $R''(0)$. Such an RG-treatment is most easily
implemented in the replica approach: The $n$ times replicated
partition function becomes after averaging over disorder
\begin{equation}\label{reps}
\overline{ \exp\!\left(-\frac{1}T \sum_{a=1}^{n}
E_{\mathrm{el}}[u_{a}] - \frac{1}T \sum_{a=1}^{n} {E_{\mathrm{DO}}
[u_{a}]} \right) } = \exp \left(-\frac1T \sum_{a=1}^{n} {
E_{\mathrm{el}}[u_{a}]} + \frac1{2T^{2}} \sum_{a,b=1}^{n} \int \rmd^d
x\, R\Big(u_{a}(x)-u_{b}(x)\Big) \right) \ .
\end{equation}
Perturbation theory is constructed along the following lines (see
\cite{BalentsDSFisher1993,LeDoussalWieseChauve2003} for more details.)  The
bare correlation function, graphically depicted as a solid line, is
with momentum $k$ flowing through and replicas $a$ and $b$
\begin{equation}
_{a}\diagram{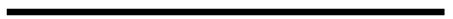}_{b}\ = \frac {T\delta_{ab}} {k^{2}}\ .
\end{equation}
The disorder vertex is
\begin{equation}
\stackrel{\!\!\!x}{\diagram{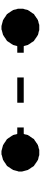}_{b}^{a}}\ =\int_{x}\sum_{a,b}
R\Big(u_{a}(x)-u_{b}(x)\Big)\ .
\end{equation}
The rules of the game are to find all contributions which correct $R$,
and which survive in the limit of $T=0$. At leading order, i.e.\ order
$R^{2}$, counting of factors of $T$ shows that only the terms with one
or two correlators contribute. On the other hand, $\sum_{a,b}
R(u_{a}-u_{b})$ has two independent sums over replicas. Thus at order
$R^{2}$ four independent sums over replicas appear, and in order to
reduce them to two, one needs at least two correlators (each
contributing a $\delta_{ab}$). Thus, at leading order, only diagrams
with two propagators survive. These are the following (noting $C(x-y)$
the Fourier transform of $1/k^{2}$):
\begin{eqnarray}\label{80}
\parbox{0mm}{\rule{0mm}{5.3mm}}^{a}_{b}\!\!\stackrel{\!\!\!x\hspace{1.4cm}y}{\diagram{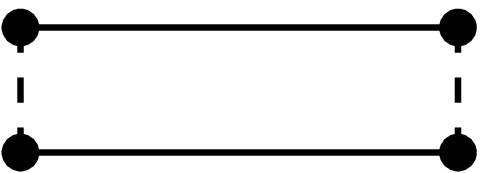}_{b}^{a}} &=& \int_{x}
 R''(u_{a}(x) -u_{b}(x))R''(u_{a}(y) -u_{a}(y))  C(x-y)^{2}
\\
\parbox{0mm}{\rule{0mm}{5.3mm}}^{a}_{b}\!\!\stackrel{\!\!\!x\hspace{1.4cm}y}{\diagram{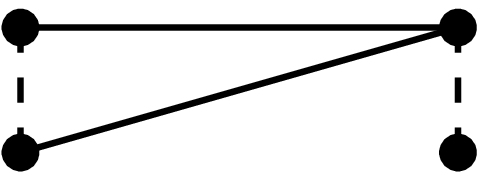}_{b}^{a}}
&=& - \int_{x} R''(u_{a}(x) -u_{a}(x))R''(u_{a}(y) -u_{a}(y))
C(x-y)^{2} \ .
\label{81}
\end{eqnarray}
In a renormalization program, we are looking for the divergences of
these diagrams. These divergences are localized at $x=y$, which allows
to approximate $R''(u_{a}(y)-u_{b}(y))$ by $R''(u_{a}(x)-u_{b}(x))$.
The integral $\int_{x-y} C(x-y)^{2} = \int_{k} \frac 1 {(
k^{2}+m^{2})^{2}} = \frac {m^{-\epsilon}}\epsilon $ (using the most
convenient normalization for $\int_{k}$) is the standard 1-loop diagram
from $\phi^{4}$-theory.  We have chosen to regulate it in the
infrared by a mass, i.e.\ physically by the harmonic well introduced in
section \ref{measurecusp}.

Note that the following diagram also contains two correlators (correct
counting in powers of temperature), but is not a 2-replica but a
3-replica sum:
\begin{equation}
{\raisebox{-1mm}{$\parbox{0mm}{\rule{0mm}{5.3mm}}^{a}_{b}$}}\!\!\stackrel{\!\!\!x\hspace{1.4cm}y}{\diagram{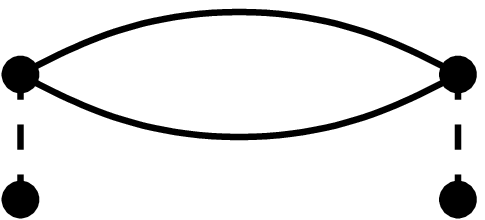}}\!\!
{\raisebox{-1mm}{$\parbox{0mm}{\rule{0mm}{5.3mm}}^{a}_{c}$}}\ \ .
\end{equation}

Taking into account combinatorial factors, the rescaling
(\ref{R-scaling}) of $R$,  as well as of the field $u$ (its dimension
being the roughness exponent $\zeta$), we arrive at
\begin{equation}\label{RG1loop}
-m\frac{\partial}{\partial m} R(u) = (\epsilon -4 \zeta) R(u) + \zeta
u R'(u) + \half R''(u)^{2} - R''(u) R''(0)\ .
\end{equation}
Note that the field does not get renormalized due to the exact
statistical tilt symmetry $u (x)\to u (x)+ \alpha x$: The bare action
(\ref{H}), including the mass term, changes according to ${\cal
H}^{\mathrm{bare}}[u] \to {\cal H}^{\mathrm{bare}}[u]+\delta {\cal
H}[u]$, with
\begin{equation}\label{a11}
\delta {\cal H}[u] = c \int\rmd^{d}x \left[ \nabla u (x) \alpha+\half
\alpha^{2} \right] + {m^{2}}\left[u (x) \alpha x +
\frac{1}{2}\alpha^{2}x^{2} \right]\ .
\end{equation}
To render the presentation clearer, the elastic constant $c$ set to
$c=1$ in equation (\ref{H}) has been introduced. The important
observation is that all fields $u$ involved are {\em large-scale}
variables, which are also present in the renormalized action, changing
according to ${\cal H}^{\mathrm{ren}}[u]\to {\cal
H}^{\mathrm{ren}}[u]+\delta H[u]$. The latter can be used to define
the renormalized elastic coefficient $c^{\mathrm{ren}}$ and mass
$m^{\mathrm{ren}}$. Since $\delta H[u]$ gives the change in energy
both for the bare and the renormalized action with unchanged
coefficients, $c^{\mathrm{ren}}\equiv c$ and $ m^{\mathrm{ren}}\equiv
m$, so neither elasticity nor mass changes under renormalization.

\section{Why is a cusp necessary in $4-\epsilon$ dimensions?}  
\begin{figure}[t] \begin{center}{\unitlength1mm\fboxsep0mm
\mbox{\begin{picture} (134,20)
\put(17,16.5){${\cal H}[u]$}
\put(52,1){$u$}
\put(98,16.5){${\cal H}[u]$}
\put(60,13.7){$L<L_{c}$}
\put(61,5.5){$L>L_{c}$}
\put(132.5,1){$u$}
\put(0,1){\fig{132mm}{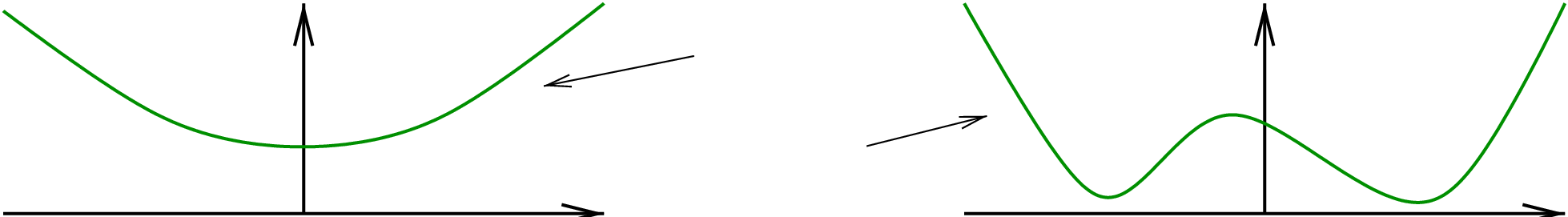} }
\end{picture} }}
\end{center}
\Caption{The toy model (\ref{toy}) before (left) and after (right) the
Larkin-scale.}\label{fig:toy}
\end{figure}

Let us
present here another simple argument due to Leon Balents
\cite{BalentsPrivate}, why a cusp {\em is a physical necessity} in
order to have a $\epsilon =4-d$ expansion. To this aim, consider a toy
model with only one Fourier-mode $u=u_{q}$
\begin{equation}\label{toy}
{\cal H}[u] = \half q^{2 } u^{2} + \sqrt{\epsilon }\, \tilde V (u)
\ .
\end{equation}
Since equation (\ref{1loopRG}) has a fixed point of order $R (u)\sim
\epsilon $ for all $\epsilon >0$, $V (u)$ scales like $\sqrt{\epsilon
}$ for $\epsilon $ small and we have made this dependence explicit in
(\ref{toy}) by using $V (u)= \sqrt{\epsilon }\tilde V (u)$. The only
further input comes from the physics: For $L<L_{c}$, i.e.\ before we
reach the Larkin length, there is only one minimum, as depicted in
figure \ref{fig:toy}. On the other hand, for $L>L_{c}$, there are
several minima. Thus there is at least one point for which
\begin{equation}
\frac{\rmd ^2}{\rmd u^2}\, {\cal H}[u] = q^{2 } + \sqrt{\epsilon }\, \tilde V''
(u) < 0
\ .
\end{equation}
In the limit of $\epsilon \to 0$, this is possible if and only if
$\frac{1}{\epsilon }R'''' (0)$, which a priori should be finite for
$\epsilon \to 0$, becomes infinite:
\begin{equation}
\frac{1}{\epsilon }R'''' (0) = \overline{V'' (u)V'' (u')}\ts _{u=u'} = \infty
\ .
\end{equation}
This argument shows that a cusp is indeed a physical necessity.

\subsection*{Acknowledgments}\label{ack} It is a pleasure to thank
the organizers of IRS-2006 for the opportunity to give this
lecture. The results presented here have been obtained in a series of
inspiring collaborations with Leon Balents, Pascal Chauve, Andrei Fedorenko, Werner
Krauth, Alan Middleton and Alberto Rosso. We also thank Dima Feldman, and Thomas
Nattermann for useful discussions.

This work has been supported by
ANR (05-BLAN-0099-01), and NSF (PHY99-07949).


\end{document}